\documentclass[twocolumn]{aastex631}
\usepackage{graphicx}
\usepackage{amstext}
\usepackage{txfonts}
\usepackage{placeins}

\usepackage{hyperref}

\hypersetup{
  colorlinks  = true, 
  urlcolor    = blue, 
  linkcolor   = blue, 
  citecolor   = blue  
}

\begin{document}

\newcommand{\arcm}{$^\prime$}
\newcommand{\arcs}{$^{\prime\prime}$}
\newcommand{\m}{$^{\rm m}\!\!.$}
\newcommand{\D}{$^{\rm d}\!\!.$}
\newcommand{\F}{$^{\rm P}\!\!.$}
\newcommand{\kms}{km~s$^{-1}$}
\newcommand{\ks}{km~s$^{-1}$}
\newcommand{\ms}{M$_{\sun}$}
\newcommand{\rs}{R$_{\sun}$}
\newcommand{\ocs}{$(O\!-\!C)/\sigma$}
\newcommand{\ubv}{\hbox{$U\!B{}V$}}
\newcommand{\ond}{Ond\v{r}ejov}
\newcommand{\jan}[1]{{\color[rgb]{0.0, 0.44, 0.0}Jan: #1}}
\newcommand{\wolf}[1]{{\color[rgb]{0.0, 0.0, 1.0}Wolf: #1}}
\newcommand{\zhar}[1]{{\color[rgb]{1.0, 0.44, 0.0}Zhar: #1}}

\title{The Z Camelopardalis-type star AY Piscium: stellar and accretion disk parameters}

\correspondingauthor{Jan K\'{a}ra}
\email{honza.kara.7@gmail.com}

   \author[0000-0002-1012-7203]{Jan K\'{a}ra}
   \affiliation{Astronomical Institute, Faculty of Mathematics and Physics, Charles University, V~Hole\v{s}ovi\v{c}k\'ach~2, CZ-180~00~Praha~8, Czech Republic}
   
 \author[0000-0003-2526-2683]{Sergey Zharikov}
 \affiliation{Universidad Nacional Aut\'{o}noma de M\'{e}xico, Instituto de Astronom\'{i}a, AP 106, Ensenada, 22800, BC, M\'{e}xico }

\author[0000-0002-4387-6358]{Marek Wolf}
 \affiliation{Astronomical Institute, Faculty of Mathematics and Physics, Charles University, V~Hole\v{s}ovi\v{c}k\'ach~2, CZ-180~00~Praha~8, Czech Republic}
 
\author [0000-0001-6964-8444] {Ainash Amantayeva }
\affiliation{Al-Farabi Kazakh National University, Al-Farabi Ave., 71, 050040, Almaty, Kazakhstan}

\author[0000-0002-0790-7292]{Gulnur Subebekova  }
\affiliation{Al-Farabi Kazakh National University, Al-Farabi Ave., 71, 050040, Almaty, Kazakhstan}

\author[0000-0001-5163-508X]{Serik Khokhlov}
\affiliation{Al-Farabi Kazakh National University, Al-Farabi Ave., 71, 050040, Almaty, Kazakhstan}

\author[0000-0001-9788-7485]{Aldiyar Agishev}
\affiliation{Al-Farabi Kazakh National University, Al-Farabi Ave., 71, 050040, Almaty, Kazakhstan}

\author[0000-0001-6355-2468]{Jaroslav Merc}
\affiliation{Astronomical Institute, Faculty of Mathematics and Physics, Charles University, V~Hole\v{s}ovi\v{c}k\'ach~2, CZ-180~00~Praha~8, Czech Republic}

 \begin{abstract}
We present a new study of the Z~Cam-type eclipsing cataclysmic variable AY~Piscium with the aim  of determining the fundamental parameters of the system and the structure of the accretion flow therein. We use time-resolved photometric observations supplemented by spectroscopy in the standstill, to which we applied our light-curve modeling techniques and the Doppler tomography method, to update system parameters. We found that the system has a massive white dwarf $M_{\rm WD}=0.90(4)$ \ms, a mass ratio  $q=0.50(3)$, and the effective temperature of a secondary $T_2 = 4100(50)$~K. The system inclination is $i=74\fdg8(7)$.  The  orbital period of the system $P_{\mathrm{orb}}=0.217320523(8)\;\mathrm{d}$ is continuously increasing with the rate of $\dot{P}_{\mathrm{orb}} = +7.6(5)\times10^{-9}$ d year$^{-1}$. The mass transfer rate varies between 2.4$\times$10$^{-10}$ M$_\odot$ year$^{-1}$ in quiescence up to 1.36$\times$10$^{-8}$ M$_\odot$ year$^{-1}$ in outburst. The accretion disk transitions from the cooler, flared, steady-state disk to a warmer state with a practically constant and relatively high disk height. The mass transfer rate is about 1.6$\times$10$^{-9}$ M$_\odot$ year$^{-1}$ in the standstill. The Balmer emission lines show a multi-component structure similar to that observed in long-orbital-period nova-like systems. Out of standstill, the system exhibits outburst bimodality, with long outbursts being more prominent. We conclude that the Balmer emission lines in AY~Psc  are  formed by the combination of radiation from the irradiated surface of the secondary, from the outflow zone, and from winds originating in the bright spot and the disk's inner part. 
\end{abstract}



\section{Introduction}

Non-magnetic cataclysmic variables (CVs) are interacting binary star systems in which a Roche-lobe–filling secondary 
(a late-type main-sequence star or a brown dwarf) transfers matter onto a white dwarf (WD) primary 
\citep{Warner:1995aa} forming an accretion disk around the WD. Typical orbital periods of CVs are between 76 minutes and 10 hours \citep{2003A&A...404..301R}. These systems are divided into several classes and subclasses generally based on their optical light 
curve behavior, e.g., Dwarf novae (DNe), Nova-likes (NLs), VY~Scl (also referred to as `anti-dwarf novae'), Z~Cam, 
and V Sge systems.
DNe systems show semi-periodic outbursts, and some of them also exhibit more energetic superoutbursts. They can be divided into several subclasses depending on the orbital period, the presence of superhumps, and the characteristic of outburst behavior (for the general information on  DNe and superhumps, see e.g., \citet{Warner:1995aa}). SU UMa-type dwarf novae ($P_{orb} \lesssim 2.2$ hours) show superhumps during their long outbursts (superoutbursts)  whose period is generally a few percent longer than the orbital period. It is generally accepted that superhumps are caused by the 3:1 resonance \citep{1988MNRAS.232...35W}. In contrast to SU UMa-type systems, DNe with longer orbital periods ($> 3$ hours)  do not show superhumps. 
Among the CVs with the shortest orbital periods are the WZ Sge-type systems, which were originally proposed as a subgroup of SU UMa-type systems with outbursts of high amplitude ($\sim 8 \, \mathrm{mag}$) and long recurrence times (up to decades) \citep{1979MNRAS.189P..41B, 2015PASJ...67..108K}.

 NLs  remain mostly  in a high luminosity state and they can be also divided into different sublcasses. The most prominent of them are SW~Sex- and VY~Scl-types stars.
 SW Sex-type systems  show distinctive spectroscopic characteristic \citep{1991AJ....102..272T} from the rest of NLs and  mostly cluster in the 3-4 h orbital period range \citep{2007MNRAS.377.1747R}. 
 The VY~Scl stars  are  a subclass of NLs whose light curves are 
 characterized by occasional drops from steady high states into low states lasting up to several hundred days. 
The Z~Cam systems are a subclass of the DNe, which in addition to outburst activity, exhibit standstills, 
during which the system luminosity stays at a high state for a long time, similarly to  NLs. The mechanisms 
of transition between outburst cycles and standstills in Z~Cam stars, and why VY~Scl stars enter a low mass transfer
state, are not well understood. It is hypothesized that VY~Scl stars might be Z~Cam stars in extended 
standstills \citep[see][section 4.1 and references therein]{Warner:1995aa}.  On the side of long-period CVs, 
there are the V~Sge-type systems \citep{1998PASP..110..276S, 1999PASP..111...76D}. The V~Sge stars' orbital periods range from 5 to 12 hours, and their orbital 
light curves are either low-amplitude sinusoidal or high-amplitude asymmetric with primary and secondary eclipses. 
Their spectra show  the ratio of the equivalent widths of \ion{He}{2}~4686 \AA\ to H$\beta$ usually greater than 
2 and indicate the presence of a strong wind in the systems \citep{1998PASP..110..276S}.

The subject of this paper is a detailed spectroscopic and photometric study of the eclipsing CV AY Piscium, hereafter AY Psc (also 
SDSS J013655.46+071629.2, 2MASS~J01365544+0716293, PG~0134+070, $V \simeq 16.3$ mag). 
 The source was identified as a CV by \citet{1982PASP...94..560G}.  \citet{1989PASP..101..899S} found that AY~Psc is 
 an eclipsing system with an orbital period of $P = 5.22$~h. Later, an orbital ephemeris was provided by \cite{1990A&A...238..170D}
 who also revised the orbital period of $P = 0.2173209\;\mathrm{d} = 5.2157\;\mathrm{h}$ . They reported 
 flickering with a semi-amplitude of about 0.08~mag present in the light curve along the whole orbital cycle. 
 
The system parameters were derived for the first time by \citet[hereafter SH93]{1993ApJ...403..743S}\defcitealias{1993ApJ...403..743S}
{SH93} using spectroscopic observations and later by \cite{1993AJ....106..311H} using photometric observations. They found that
the system contains a massive $\sim$ 1.11-1.31~\ms\ white dwarf, a $\sim$ 0.53-0.59~\ms\ companion, and that the system inclination
is about 74$^\circ$. Those estimates locate  AY~Psc in a harbor of potential SN~Ia progenitors.

\cite{2002AAS...201.4007M} classified AY~Psc as a Z~Cam-type star based on the presence of dwarf-nova outbursts
as well as occasional standstills.
\cite{2009NewA...14..330G} reported an extensive photometric analysis of AY~Psc based on observation obtained during 
a large number of nights in 2003--2005. They confirmed the stability of the system's orbital period in 
the long term and found evidence of variable negative superhumps in the system. \cite{2017RAA....17...56H} continued to 
study the long-term photometric behaviour of AY~Psc. They reported that the system shows outbursts with an amplitude of $\sim 2.5$ mag, 
during which the system reaches the maximal brightness $V\simeq 14.6$ mag.  Outbursts repeated themselves with a period 
of $\sim 18.3(7)$~days.

The first time-resolved spectroscopy of AY~Psc was reported by \citetalias{1993ApJ...403..743S}. The object had 
15.8 mag in the V-band, which is close to the low state level. They covered about 70 \% of an orbit in the wavelength range 
of 4250-4950 \AA\ with a spectral resolution of about 2.8 \AA. The presented spectra show complex Balmer line profiles 
and the absence of the \ion{He}{2}~$\lambda$4686 line. The full width zero intensity (FWZI) of H$\beta$ was between 30 and 40~\AA. 
No other time-resolved spectroscopy was reported in the literature. 

Three additional spectra were acquired by the LAMOST 
project  \citep{2020PASJ...72...76H}. The spectra were obtained close to the high brightness state ($V_{\rm AAVSO}\simeq 
14.6$~mag). They are significantly different compared to the low state spectra of \citetalias{1993ApJ...403..743S} and
exhibit broad single-peaked Balmer emission lines and several weak neutral helium emission lines (e.g., HeI $\lambda$4914, 
$\lambda$5876, $\lambda$6678). There are  also strong \ion{He}{2} $\lambda$4686 and Bowen blend lines 
(\ion{C}{3}/\ion{N}{3} $\lambda$4634-4651).  Some absorption features at \ion{He}{1} $\lambda$4026 and $\lambda$4471, 
\ion{Fe}{1} $\lambda$5018, \ion{Fe}{2} $\lambda$4924, \ion{Mg}{1}$~\lambda$5178 and Na D lines were also found. 
The FWZI and equivalent width (EW) of H$\beta$ and  H$\alpha$ lines in LAMOST spectra were 42.1\AA~/5.0\AA\ 
and 120.7\AA~/17.6\AA, respectively.

 The {\sc Gaia} parallax  
of AY~Psc from DR3 \citep{2022yCat.1355....0G} is $\pi = 1.37(4)$ mas,
which gives a distance of $d =$ 730(20) pc.
The interstellar absorption in the direction of AY Psc is $E(B-V) = 0.05\pm 0.03$ \citep{2011ApJ...737..103S}, three-dimensional map of interstellar dust by \cite{2015ApJ...810...25G} gives for the same direction and for the object's distance absorption of $E(g-r) = 0.04^{+0.05}_{-0.01}$.

AY~Psc is the only confirmed Z~Cam-type DN that
shows deep eclipses (0.75-1.75  mag)\footnote{EM Cyg, another Z~Cam-type system, exhibits eclipses of depth about 0.15 mag \citep{2021MNRAS.505..677L} caused by the system inclination of $i=67\pm2$ \citep{2000MNRAS.313..383N} at which only a small part of the accretion disk is obscured by the secondary.}, which provides an opportunity for a thorough analysis of a system belonging to this subclass of DNe.
The aims of the presented paper are to determine the fundamental parameters of the system based on the analyses of the light curves obtained in different states of the system, study the structure of the accretion flow therein, and the origin of the sources that form emission line profiles. 
This will be useful to better understand the mechanism of switching between outburst activity and standstills in Z~Cam systems.

\section{Observations}
\label{sec:obs}

\begin{figure*}[hbt]
   \centering
    \includegraphics{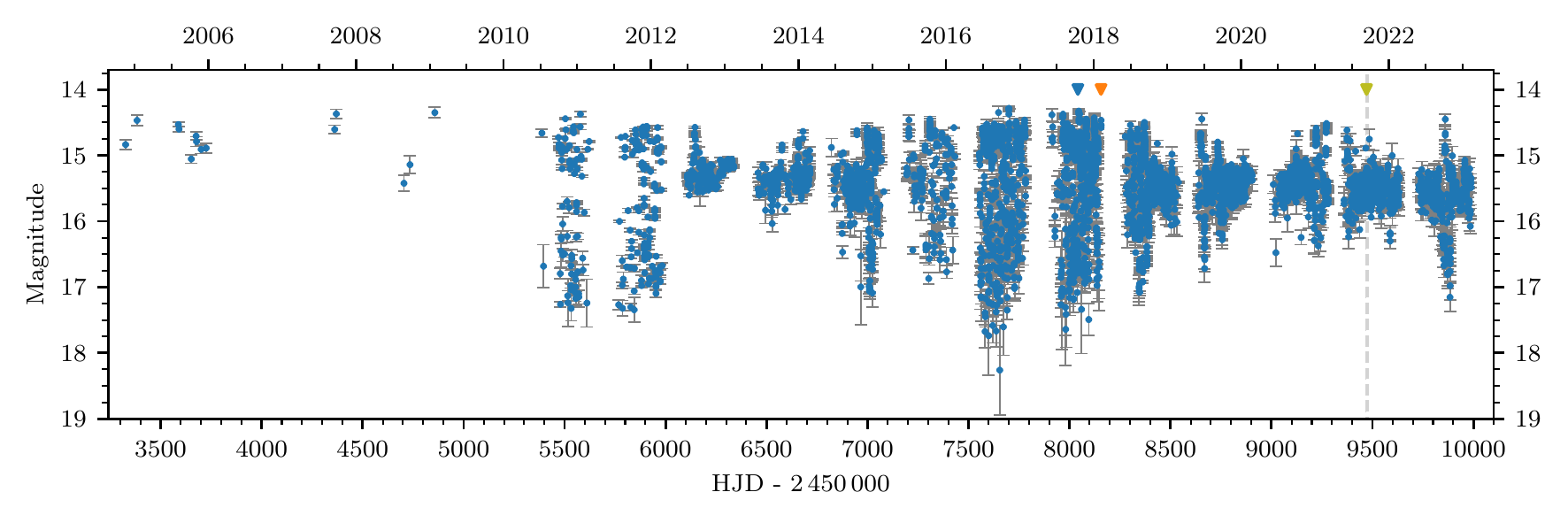}
    \includegraphics{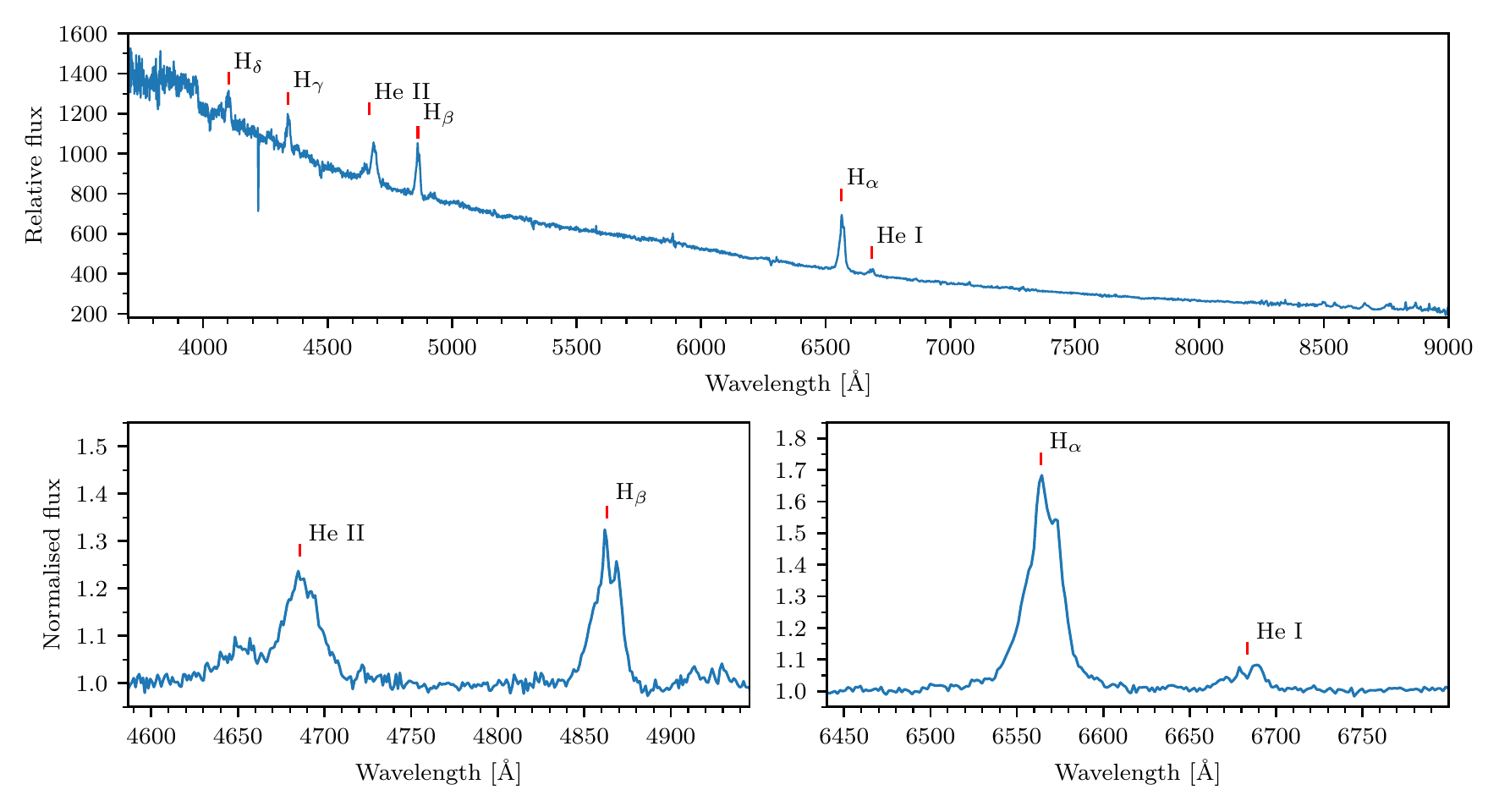}
   \caption{{\it Top:}  Long-term light curve of AY~Psc collected by AAVSO, ASAS, ASAS-SN, ZTF, and our photometry obtained at OAN SPM, \ond, La Silla and TSHAO, only out-of-eclipse measurements are shown. Grey dashed lines mark the times of the spectroscopic observations obtained at OAN SPM. The triangles mark the nights when photometric observations used for light-curve modeling were obtained, their color corresponds to the eclipse light curves discussed below. These light curves were obtained at HJDs $2\,458\,041$, $2\,459\,470$, and $2\,458\,156$ and show the light curve of AY~Psc in quiescence, outburst, and standstill, respectively.
   {\it Middle and bottom panels:}  LAMOST spectrum of AY~Psc (middle panel) and cutouts of normalized LAMOST spectrum centered on H${\beta}$ and \ion{He}{2} {\it (bottom, left)}, and H${\alpha}$ and \ion{He}{1} {\it (bottom, right).}  The air wavelength positions of the emission lines are marked by red lines. }
   
 \label{F:01}
\end{figure*}

\begin{table*}[!tbp]
\caption{Log of observations}
\label{T:LOG}
\centering
\begin{tabular}{l l c c c l }
			\hline \hline										
	\multicolumn{6}{c}{		\textbf{		Photometry		}}					\\
	\hline 										
	HJD	&	Civil date	&	Filter	&	Exposure time	&	Number of frames	&	Observatory	\\
$	-2\,400\,000	$&		&		&	[s]	&		&		\\
	\hline 										
$	58\,041	$&	2017 Oct 14	&	C	&$	60	$&$	91	$&	\ond	\\
$	58\,156	$&	2018 Feb 6	&	C	&$	60	$&$	50	$&	\ond	\\
$	58\,847	$&	2019 Dec 29	&	C	&$	60	$&$	48	$&	\ond	\\
$	59\,184	$&	2020 Nov 30	&	R	&$	30	$&$	57	$&	La Silla	\\
$	59\,471	$&	2021 Sep 13	&	V	&$	60	$&$	308	$&	OAN SPM	\\
$	59\,472	$&	2021 Sep 14	&	V	&$	60	$&$	296	$&	OAN SPM	\\
$	59\,473	$&	2021 Sep 15	&	V	&$	60	$&$	350	$&	OAN SPM	\\
$	59\,474	$&	2021 Sep 16	&	V	&$	60	$&$	314	$&	OAN SPM	\\
$	59\,494	$&	2021 Oct 6	&	V	&$	60	$&$	386	$&	OAN SPM	\\
$	59\,497	$&	2021 Oct 9	&	V	&$	60	$&$	244	$&	OAN SPM	\\
$	59\,498	$&	2021 Oct 10	&	V	&$	60	$&$	240	$&	OAN SPM	\\
$	59\,511	$&	2021 Oct 23	&	C	&$	60	$&$	54	$&	\ond	\\
$	59\,515	$&	2021 Oct 27	&	C	&$	60	$&$	44	$&	\ond	\\
$	59\,528	$&	2021 Nov 9	&	V	&$	60	$&$	105	$&	TSHAO	\\
$	59\,571	$&	2021 Dec 22	&	C	&$	60	$&$	49	$&	\ond	\\
$	59\,586	$&	2022 Jan 6	&	C	&$	60	$&$	32	$&	\ond	\\
$	59\,857	$&	2022 Oct 4	&	C	&$	60	$&$	45	$&	\ond	\\
$	59\,874	$&	2022 Oct 21	&	C	&$	60	$&$	33	$&	\ond	\\
$	59\,898	$&	2022 Nov 14	&	C	&$	60	$&$	46	$&	\ond	\\
	\hline  									
	\multicolumn{6}{c}{		\textbf{		Spectroscopy		}}					\\
	\hline   									
	HJD	&	Civil date	&	Spectral range	&	Exposure time	&	Number of Spectra	&	Observatory	\\
$	-2\,400\,000	$&		&		&	[s]	&		&		\\
	\hline   									
$	59\,471	$&	2021 Sep 13	&	6026 \AA - 7258 \AA	&$	1\,200	$&$	15	$&	OAN SPM	\\
$	59\,472	$&	2021 Sep 14	&	6026 \AA - 7258 \AA	&$	1\,200	$&$	15	$&	OAN SPM	\\
$	59\,473	$&	2021 Sep 15	&	6026 \AA - 7258 \AA	&$	1\,200	$&$	15	$&	OAN SPM	\\
$	59\,474	$&	2021 Sep 16	&	6026 \AA - 7258 \AA	&$	1\,200	$&$	15	$&	OAN SPM	\\
	\hline 																			
\end{tabular}

\end{table*}

Over the past years, we have accumulated new AY~Psc data, including our own photometric and spectroscopic measurements obtained at several observatories as well as available data or information previously published in the literature. A log of the newly obtained observations is given in Table~\ref{T:LOG}.

\subsection{Photometry}

\begin{table}[!tbp]
\caption{New mid-eclipse times of AY~Psc determined from ground-based observations}  
\label{Tab:ecl}
\begin{center}
\begin{tabular}{lllcc}
\hline\noalign{\smallskip}  
 $\mathrm{BJD}_{\mathrm{min}}$ -- &  Epoch & Error  & Filter & Observatory \\
2\,400\,000    &        & [day]  &        &            \\ 
\hline\noalign{\smallskip}  
52\,966.17334 &  24\,585 & 0.0005 & C & Exmouth \\
52\,968.12946 &  24\,594 & 0.0005 & C & Exmouth \\
52\,969.21630 &  24\,599 & 0.0005 & C & Exmouth \\
52\,972.04142 &  24\,612 & 0.0005 & C & Exmouth \\
52\,986.16834 &  24\,677 & 0.0005 & C & Exmouth \\
52\,987.03754 &  24\,681 & 0.0005 & C & Exmouth \\
52\,988.12446 &  24\,686 & 0.0005 & C & Exmouth \\
53\,272.15809 &  25\,993 & 0.0005 & C & Exmouth \\
53\,273.24509 &  25\,998 & 0.0005 & C & Exmouth \\
53\,287.15469 &  26\,062 & 0.0005 & C & Exmouth \\
53\,291.06605 &  26\,080 & 0.0005 & C & Exmouth \\
53\,291.28355 &  26\,081 & 0.0005 & C & Exmouth \\
53\,292.15251 &  26\,085 & 0.0005 & C & Exmouth \\
53\,294.10878 &  26\,094 & 0.0005 & C & Exmouth \\
53\,297.15130 &  26\,108 & 0.0005 & C & Exmouth \\
53\,663.12073 &  27\,792 & 0.0005 & C & Exmouth \\
53\,670.29257 &  27\,825 & 0.0005 & C & Exmouth \\
53\,675.07403 &  27\,847 & 0.0005 & C & Exmouth \\
53\,675.29136 &  27\,848 & 0.0005 & C & Exmouth \\
53\,676.16109 &  27\,852 & 0.0005 & C & Exmouth \\
53\,677.24797 &  27\,857 & 0.0005 & C & Exmouth \\
53\,678.11698 &  27\,861 & 0.0005 & C & Exmouth \\
58\,041.48048 &  47\,939 & 0.0001 & C & \ond \\
58\,156.22572 &  48\,467 & 0.0001 & C & \ond \\
58\,778.85019 &  51\,332 & 0.0005 & r & ZTF \\
58\,779.93694 &  51\,337 & 0.0005 & r & ZTF \\
58\,847.30611 &  51\,647 & 0.0001 & C & \ond \\
59\,184.58819 &  53\,199 & 0.0001 & R & La Silla \\
59\,470.79944 &  54\,516 & 0.0001 & V & OAN SPM \\ 
59\,471.01677 &  54\,517 & 0.0001 & V & OAN SPM \\ 
59\,471.88628 &  54\,521 & 0.0001 & V & OAN SPM \\ 
59\,472.97279 &  54\,526 & 0.0001 & V & OAN SPM \\
59\,473.84180 &  54\,530 & 0.0001 & V & OAN SPM \\ 
59\,493.83522 &  54\,622 & 0.0001 & V & OAN SPM \\ 
59\,496.87772 &  54\,636 & 0.0001 & V & OAN SPM \\
59\,497.96454 &  54\,641 & 0.0001 & V & OAN SPM \\
59\,511.43831 &  54\,703 & 0.0001 & C & \ond \\
59\,515.34967 &  54\,721 & 0.0002 & C & \ond \\
59\,528.17230 &  54\,780 & 0.0001 & V & TSHAO \\
59\,571.20157 &  54\,978 & 0.0001 & C & \ond \\
59\,586.19618 &  55\,047 & 0.0002 & C & \ond \\
 59\,857.41284 & 56\,295 & 0.0001 & C & \ond \\
 59\,898.26921 & 56\,483 & 0.0001 & C & \ond \\
\hline
\end{tabular} 
\end{center}
\tablecomments{
 The letters in the "Filter" column mark: "V" -- Bessel V-band, "R" -- Bessel R-band, "C" -- clear, observations without filter, "r" -- ZTF-r filter, respectively.}

\end{table}

New ground-based photometry\footnote{ All data were reduced  using  the {\it apphot} IRAF task except \ond\ Observatory and La Silla Observatory. The heliocentric correction was applied. The magnitude errors, ranging from 0.01 to 0.02~mag (OAN SPM, La Silla, and TSHAO) and 0.01-0.05~mag (\ond) were estimated from the magnitude dispersion of comparison stars with similar brightness.} of AY~Psc was obtained at four different observatories:
Observatorio Astronomico Nacional San Pedro M\'artir (Mexico, hereafter OAN SPM), 
\ond\ Observatory (Czech Republic, hereafter \ond),
La Silla Observatory in Chile, and 
Tien Shan Astronomical Observatory (Kazakhstan, hereafter TSHAO). 
Below we describe the details of these observations  shortly for each observatory.
 These newly obtained data, together with recently available ones, were used for new mid-eclipse times determination presented in Table~\ref{Tab:ecl}. 
New times of primary eclipses and their uncertainties were generally
determined by fitting the light curve by Gaussians or polynomials of
the third or fourth order. The least-squares method to the lower part 
of the light curve  was applied. The times were converted from heliocentric Julian date (HJD) to barycentric Julian date (BJD) using a conversion tool created by  \cite{2010PASP..122..935E}\footnote{Available at \url{https://astroutils.astronomy.osu.edu/time/hjd2bjd.html}}.

\begin{table*}[!tbp]

\caption{Mid-eclipse times of AY~Psc based on TESS observations}  
\label{Tab:ecl_TESS}
\centering
\begin{tabular}{ll | ll | ll | ll }
\hline\noalign{\smallskip}  
 $\mathrm{BJD}_{\mathrm{min}}$ -- &  Epoch  & $\mathrm{BJD}_{\mathrm{min}}$ -- &  Epoch & $\mathrm{BJD}_{\mathrm{min}}$ -- &  Epoch  & $\mathrm{BJD}_{\mathrm{min}}$ -- &  Epoch  \\
2\,400\,000    &        &  2\,400\,000 &    &  2\,400\,000    &        &2\,400\,000    &        \\
\hline\noalign{\smallskip}  
$ 59\,447.76307   $ & $ 54\,410   $     &   $ 59\,453.19610   $ & $ 54\,435   $    &   $ 59\,461.67160   $ & $ 54\,474   $    &   $ 59\,467.10444   $ & $ 54\,499   $ \\
$ 59\,447.98041   $ & $ 54\,411   $     &   $ 59\,453.41342   $ & $ 54\,436   $    &   $ 59\,461.88885   $ & $ 54\,475   $    &   $ 59\,467.32195   $ & $ 54\,500   $ \\
$ 59\,448.19767   $ & $ 54\,412   $     &   $ 59\,453.63074   $ & $ 54\,437   $    &   $ 59\,462.10625   $ & $ 54\,476   $    &   $ 59\,467.53927   $ & $ 54\,501   $ \\
$ 59\,448.41506   $ & $ 54\,413   $     &   $ 59\,453.84804   $ & $ 54\,438   $    &   $ 59\,462.32357   $ & $ 54\,477   $    &   $ 59\,467.75658   $ & $ 54\,502   $ \\
$ 59\,448.63237   $ & $ 54\,414   $     &   $ 59\,454.06539   $ & $ 54\,439   $    &   $ 59\,462.54087   $ & $ 54\,478   $    &   $ 59\,467.97388   $ & $ 54\,503   $ \\
$ 59\,448.84974   $ & $ 54\,415   $     &   $ 59\,454.28269   $ & $ 54\,440   $    &   $ 59\,462.75841   $ & $ 54\,479   $    &   $ 59\,468.19123   $ & $ 54\,504   $ \\
$ 59\,449.06702   $ & $ 54\,416   $     &   $ 59\,454.50002   $ & $ 54\,441   $    &   $ 59\,462.97553   $ & $ 54\,480   $    &   $ 59\,468.40855   $ & $ 54\,505   $ \\
$ 59\,449.28440   $ & $ 54\,417   $     &   $ 59\,454.71735   $ & $ 54\,442   $    &   $ 59\,463.19286   $ & $ 54\,481   $    &   $ 59\,468.62585   $ & $ 54\,506   $ \\
$ 59\,449.50173   $ & $ 54\,418   $     &   $ 59\,454.93467   $ & $ 54\,443   $    &   $ 59\,463.41019   $ & $ 54\,482   $    &   $ 59\,468.84319   $ & $ 54\,507   $ \\
$ 59\,449.71902   $ & $ 54\,419   $     &   $ 59\,455.15189   $ & $ 54\,444   $    &   $ 59\,463.62749   $ & $ 54\,483   $    &   $ 59\,469.06051   $ & $ 54\,508   $ \\
$ 59\,449.93629   $ & $ 54\,420   $     &   $ 59\,455.36932   $ & $ 54\,445   $    &   $ 59\,463.84483   $ & $ 54\,484   $    &   $ 59\,469.27774   $ & $ 54\,509   $ \\
$ 59\,450.15360   $ & $ 54\,421   $     &   $ 59\,455.58664   $ & $ 54\,446   $    &   $ 59\,464.06213   $ & $ 54\,485   $    &   $ 59\,469.49525   $ & $ 54\,510   $ \\
$ 59\,450.37090   $ & $ 54\,422   $     &   $ 59\,455.80384   $ & $ 54\,447   $    &   $ 59\,464.27937   $ & $ 54\,486   $    &   $ 59\,469.71252   $ & $ 54\,511   $ \\
$ 59\,450.58833   $ & $ 54\,423   $     &   $ 59\,456.02126   $ & $ 54\,448   $    &   $ 59\,464.49680   $ & $ 54\,487   $    &   $ 59\,469.92979   $ & $ 54\,512   $ \\
$ 59\,450.80557   $ & $ 54\,424   $     &   $ 59\,456.45593   $ & $ 54\,450   $    &   $ 59\,464.71416   $ & $ 54\,488   $    &   $ 59\,470.14719   $ & $ 54\,513   $ \\
$ 59\,451.02299   $ & $ 54\,425   $     &   $ 59\,456.89053   $ & $ 54\,452   $    &   $ 59\,464.93149   $ & $ 54\,489   $    &   $ 59\,470.36450   $ & $ 54\,514   $ \\
$ 59\,451.24022   $ & $ 54\,426   $     &   $ 59\,457.10794   $ & $ 54\,453   $    &   $ 59\,465.14873   $ & $ 54\,490   $    &   $ 59\,470.58199   $ & $ 54\,515   $ \\
$ 59\,451.45753   $ & $ 54\,427   $     &   $ 59\,457.54246   $ & $ 54\,455   $    &   $ 59\,465.36606   $ & $ 54\,491   $    &   $ 59\,470.79916   $ & $ 54\,516   $ \\
$ 59\,451.67482   $ & $ 54\,428   $     &   $ 59\,458.41193   $ & $ 54\,459   $    &   $ 59\,465.58337   $ & $ 54\,492   $    &   $ 59\,471.01640   $ & $ 54\,517   $ \\
$ 59\,451.89218   $ & $ 54\,429   $     &   $ 59\,458.62915   $ & $ 54\,460   $    &   $ 59\,465.80070   $ & $ 54\,493   $    &   $ 59\,471.23359   $ & $ 54\,518   $ \\
$ 59\,452.10962   $ & $ 54\,430   $     &   $ 59\,458.84644   $ & $ 54\,461   $    &   $ 59\,466.01802   $ & $ 54\,494   $    &   $ 59\,471.45103   $ & $ 54\,519   $ \\
$ 59\,452.32682   $ & $ 54\,431   $     &   $ 59\,459.06377   $ & $ 54\,462   $    &   $ 59\,466.23533   $ & $ 54\,495   $    &   $ 59\,471.66836   $ & $ 54\,520   $ \\
$ 59\,452.54415   $ & $ 54\,432   $     &   $ 59\,459.28105   $ & $ 54\,463   $    &   $ 59\,466.45264   $ & $ 54\,496   $    &   $ 59\,472.53763   $ & $ 54\,524   $ \\
$ 59\,452.76147   $ & $ 54\,433   $     &   $ 59\,460.15036   $ & $ 54\,467   $    &   $ 59\,466.67006   $ & $ 54\,497   $    &   $ 59\,472.75504   $ & $ 54\,525   $ \\
$ 59\,452.97879   $ & $ 54\,434   $    &   $ 59\,461.45431   $ & $ 54\,473   $    &   $ 59\,466.88728   $ & $ 54\,498   $    &   $ $ & $ $  \\
\hline
\end{tabular}

\smallskip
Note. We assumed accuracy of $0.0005\;\mathrm{d}$ for all mid-eclipse times determined from the light curve observed by TESS.
\end{table*}

\subsubsection{OAN SPM}

Time-resolved photometry of AY~Psc was obtained using the direct CCD image mode of the 0.84-m telescope of OAN SPM. The object was observed repeatedly during  seven photometric nights on October 13--16 and November 6--10, 2021. The time of individual exposure in the V-band was 60 seconds, and each run lasted about 5--8~hours. The images were bias-corrected and flat-fielded before aperture photometry was carried out.The photometric data were calibrated using the field standard stars  from \citet{1995PASP..107..324H}.

\begin{figure}[t]
\includegraphics{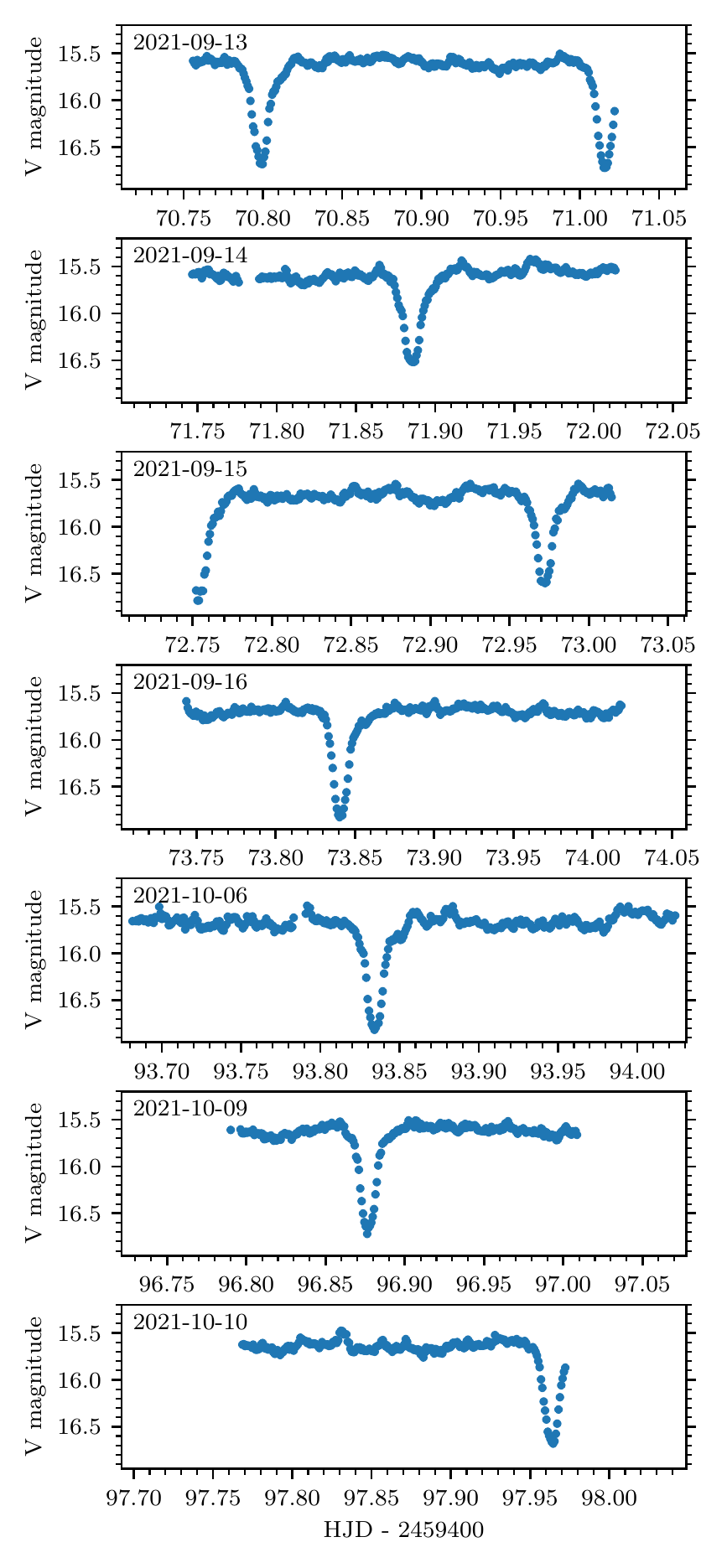}
\caption{Light curves of AY~Psc showing primary eclipses obtained at OAN SPM during 7 nights in October and November 2021, all light curves were obtained during standstill}. 
\label{fig:EclipseLC:SPM}
\end{figure}

\begin{figure*}[t]
\includegraphics{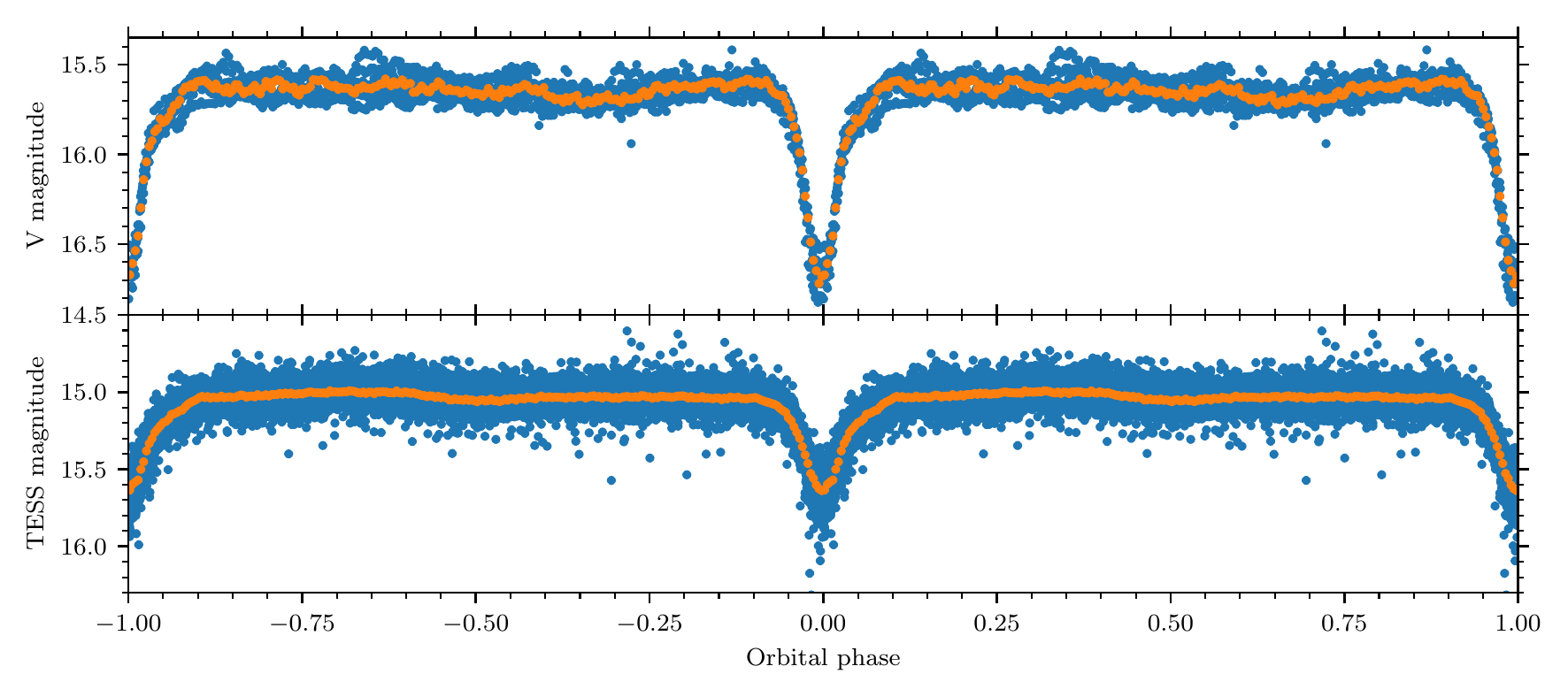}
\caption{ Phase-folded light curve of AY~Psc in V-band at OAN SPM ({\it top}) and TESS magnitude ({\it bottom}), both light curves were obtained during standstill. The light curves were phase-folded orbital $P_{\mathrm{orb}} = 0.21732002\;\mathrm{d}$, the phase $\varphi = 0.0$ corresponds to the inferior conjunction of the secondary. The out-of-eclipse TESS magnitude was estimated using the {\sc ticgen} software package \citep{tom_barclay_2017_888217, 2018AJ....156..102S} with out-of-eclipse brightness in V filter as an input parameter. The blue circles mark the individual measurements, the orange circles represent the phase-averaged light curve computed as median values in evenly-spaced bins with a width of $\Delta \varphi =  0.004$ orbital phase. 
}
\label{fig:TESS_MEAN}
\end{figure*}

\subsubsection{\ond}

Time-resolved CCD photometry of AY~Psc has also been obtained 
at the \ond\ Observatory in the years 2017--2022. The Mayer 
0.65-m ($f/3.6$) reflecting telescope with the CCD camera MI G2-3200 without a photometric filter (marked by "C" in Tables~\ref{T:LOG} and \ref{Tab:ecl}) was used. The {\sc Aphot}\footnote{Developed at the \ond\ Observatory by M. Velen and P. Pravec.}, synthetic aperture photometry and astrometry software, was used for data reduction. The quality of the nights was photometric or with light clouds.  Therefore, differential photometry was performed using suitable comparison stars also from \citet{1995PASP..107..324H}.

\subsubsection{La Silla and TSHAO}

Two additional time series were obtained at La Silla and TSHAO observatories.
In the first case, the Danish 1.54-m telescope was used in remote access on December 1, 2020. The R filter and 30-second exposure time were used, and the conditions were photometric. The {\sc Aphot} software was used for data reduction.

In TSHAO, the photometric data were obtained using  the 1-m telescope on November 9, 2021, in  the V- band and with 60 seconds exposures. The observations were obtained under photometric weather conditions.

\subsubsection{Exmouth}

The original unfiltered CCD photometry obtained by A.~Liu at Norcape Observatory, Exmouth, Western Australia, during the 2003--2005 seasons 
(48~nights together) and used primarily for search in a superhump period of AY~Psc \citep{2009NewA...14..330G} was kindly provided to us by H.~G{\"u}lsecen, Istanbul University. 

\subsubsection{ASAS, ASAS-SN, ZTF, AAVSO} 

We have also employed photometric measurements of AY~Psc obtained by the All-Sky Automated Survey\footnote{\url{http://www.astrouw.edu.pl/asas/}} \citep[ASAS; $V$ filter;][]{1997AcA....47..467P}, All-Sky Automated Survey for Supernovae\footnote{\url{https://asas-sn.osu.edu/}} \citep[ASAS-SN; \textit{V}~and \textit{g} filters; ][]{2014ApJ...788...48S, 2017PASP..129j4502K}, the Zwicky Transient Facility (ZTF) survey\footnote{\url{https://www.ztf.caltech.edu/ztf-public-releases.html} \citep{https://doi.org/10.26131/irsa539}} \citep[\textit{r}~and \textit{g} filters; ][]{2019PASP..131a8003M} and the database of American Association of Variable Star Observers\footnote{\url{https://www.aavso.org}} \citep[AAVSO; \textit{V} filter;][]{AAVSO:ONLINE}. The photometric data were subjected to visual inspection and used for the construction of the long-term light curve. 
Using the data from the ZTF survey, we derived two additional mid-eclipse times, and they are also given in Table~\ref{Tab:ecl}.

\subsubsection{TESS}
  AY~Psc was also measured by the {\sc TESS} satellite  in a 2-minute cadence during two consecutive periods (Sectors 42 and 43) in August and September 2021 during a standstill. We also used these data\footnote {The TESS data used in this paper can be found in MAST: \dataset[https://doi.org/10.17909/5fq3-s930]{https://doi.org/10.17909/5fq3-s930}.} for mid-eclipse time determination. However, the poor coverage of the light curve during eclipses reduced the achievable precision, a fact which was taken into account in further analysis, where we assumed an accuracy of $0.0005$ d for all times of eclipses. The times of eclipses and corresponding epochs are given in Table~\ref{Tab:ecl_TESS}.


\subsection{Spectroscopy}

We obtained new time-resolved spectroscopic observations of the H$\alpha$ line. All new observations were obtained during a standstill. We also used the spectra obtained by the LAMOST project\footnote{\url{http://www.lamost.org/dr8/v2.0/}  (R=1800 at $\lambda = 5500$ \AA)} \citep{2020PASJ...72...76H}, which are shown in Fig.\ref{F:01}, middle and bottom panels.

\subsubsection{Time-resolved H$\alpha$ spectroscopy}

The new spectroscopic data of AY~Psc were obtained using  
the Boller \& Chivens long-slit spectrograph attached 
to the 2.12-m telescope of the OAN SPM\footnote{\url{http://www.astrossp.unam.mx}}.  
A total of 60 spectra were obtained 
during four consecutive nights, log of the observations is listed in Table~\ref{T:LOG}. 
The Boller \& Chivens spectrograph covered spectral range 6026--7258~\AA\ with 
the spectral resolving power of $R \approx$ 5500. We obtained 15 spectra with exposure time of $1\,200\;\mathrm{s}$ each night, the duration of each observing run was $4.9\;\mathrm{h}$.  
The weather conditions during the observing run were photometric. They were controlled by simultaneous photometric observations of the object.
The spectral data were reduced in the standard way using the {\it apextract} and {\it onedspec} IRAF tools.
Standard procedures, including bias and flat-field correction, cosmic ray removal, and wavelength calibration, were applied.

\section{The AY~Psc light curve analysis}
\label{sec:lc}

All long-term photometry of the AY~Psc collected at different sites is presented in Fig.~\ref{F:01}.  The light curve clearly demonstrates the periods of DNe activity of the system and the time/duration of standstills.
Individual V-band time-series OAN~SPM  data  obtained in 2021 during standstills are given in Fig.~\ref{fig:EclipseLC:SPM}. Flickering with an amplitude of about 0.15 mag is well-visible in out-of-eclipse phases of the light curve. The depth of the eclipse of the disk by the secondary was about 1.10(15) mag.
In contrast with the absence of the bright spot contribution in the photometric data reported by \citetalias{1993ApJ...403..743S}, 
we clearly see some  hints of it in our data (see the light curve before the eclipse in Fig.~\ref{fig:EclipseLC:SPM}, HJD-2459400: 72.9--73.0 and 96.8--96.9), and in some light curves from \citet{2009NewA...14..330G}.   The light curve in standstill folded on the orbital period is presented in the top panel of  Fig.~\ref{fig:TESS_MEAN} in comparison with similar TESS data. The blue points show raw data, and the orange points show the average light curve, which was computed as median values in evenly-spaced bins with a width of $\Delta \varphi = 0.004$. 
The depth of the main eclipse in the TESS  data is about 0.7(3)~mag. It is less than in the V band, as the spectral range of TESS is centered on longer wavelengths. There is also a faint hint of a secondary eclipse  ($\varphi = 0.5$)
in the smoothed TESS light curve. 

\subsection{Long-term variability}
\label{S:LTV}

\begin{figure*}
   \centering
   \includegraphics{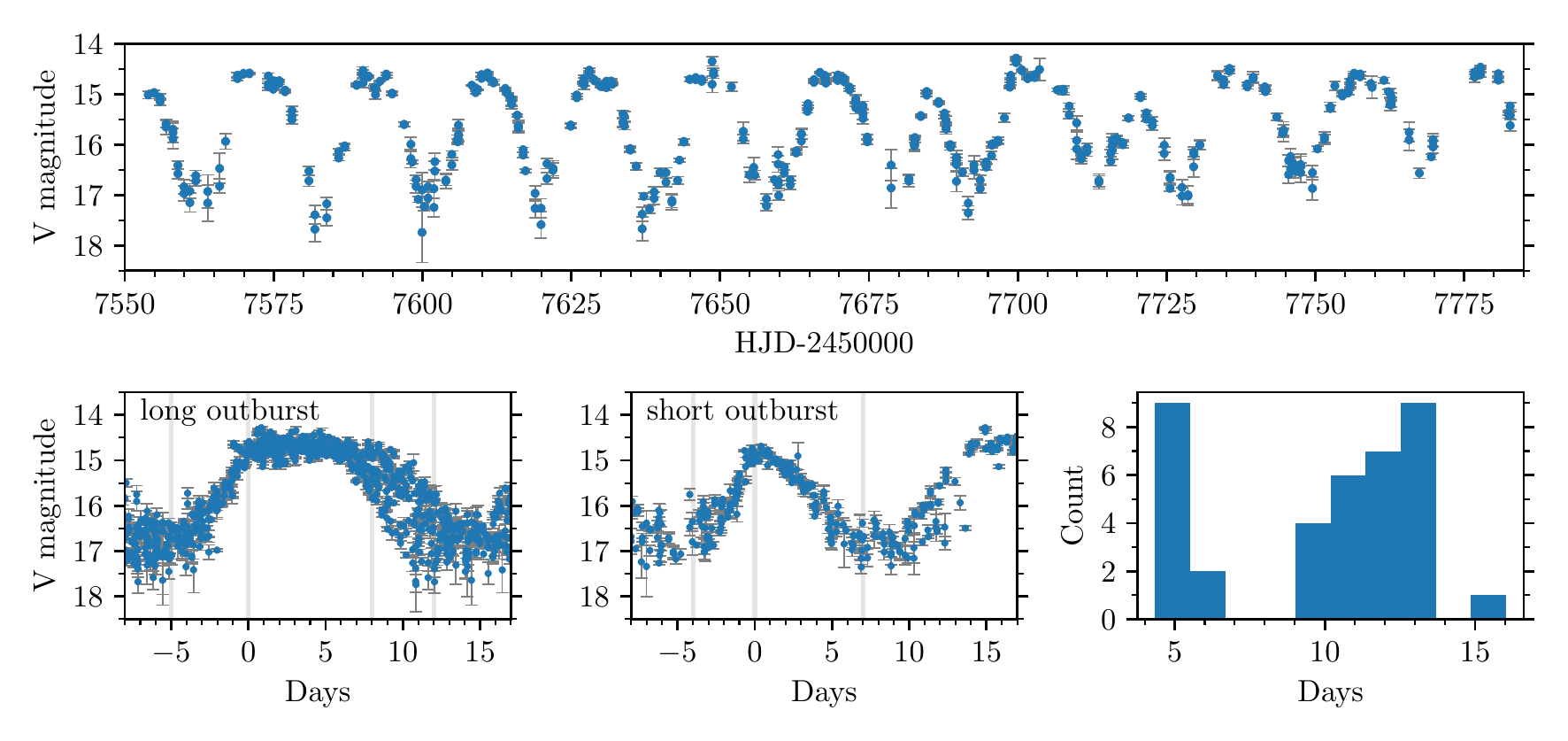}
   \caption{Example of an outbursting phase as covered by ASAS-SN an AAVSO data ({\it top}), only out-of-eclipse measurements are shown, average light curves of long  and short outbursts ({\it bottom left and center}, respectively) and histogram showing the distribution of FWHM of analyzed outbursts ({\it bottom right}). The average light curves were constructed using ASAS-SN and AAVSO data obtained out fo eclipse, which were shifted by an arbitrary time value so that the shapes of the outbursts are outlined. The average light curves consist of 27 and 11 separate outburst light curves for long and short outbursts, respectively.  The grey vertical lines show our definition of moments of the beginning of outbursts, reaching a maximal brightness,  the end of the plateau phase of the long outbursts, and the moment when the disk returns to quiescence.}
 \label{F:ON}
\end{figure*}

Examples of outburst behavior are shown in Fig.~\ref{F:ON}, where in the top panel, we present outbursts detected in the time interval between HJD~2\,457\,550--2\,457\,775.
 Based on their shape and duration, the outbursts can be divided into two different classes designated as 'long' and 'short' outbursts.
 This bimodality in outburst duration is present in many DNe \citep{1984PASP...96..988S}, most prominently 
 in SU~UMa DNe, which exhibit outbursts and superoutbursts \citep{2016MNRAS.460.2526O}.
 The average light curves of both classes are presented in the lower panels of Fig.~\ref{F:ON}. The light curves were formed by a combination of data of outbursts obtained from the AAVSO and ASAS-SN databases, which were shifted by an arbitrary time value so that the shape of an outburst is outlined. The time when the outburst reaches its maximal brightness was chosen as the reference point and set as time zero. 
We analyzed 38 outbursts observed during active states for which good observational coverage was found in AAVSO and ASAS-SN databases.
We classified 27 of them (71\%) as long outbursts and 11 (29\%) as short outbursts. The right-bottom panels of Fig.~\ref{F:ON} show 
the distribution of full width at half maximum (FWHM) of outbursts, and  in Table~\ref{Tab:Out}, we summarised their main characteristics. 
Most of the observed outbursts were long outbursts, which is in contrast with most other Z~Cam DNe, for which short outbursts are more common \citep{1979JBAA...89..169C, 1984PASP...96..988S, 1998AJ....115.1175O}, and also with other DNe exhibiting long and short outbursts \citep[see fig.~12 therein]{2016MNRAS.460.2526O}.

The amplitudes of outbursts are $A_{\mathrm{O}} \simeq $ 2.5~mag for the long outbursts and $A_{\mathrm{O}} \simeq $ 2.0~mag for the short outbursts.  Typically, the system reaches the maximal brightness during an outburst in 4--5 days. 
In the case of long outbursts, the system remains in a plateau phase for several days (from 6 to 9) before the brightness starts to decrease. 
In the case of short outbursts, the decrease in brightness follows right after reaching their maximum, and the plateau phase is absent. 
The decrease rate of short outbursts is stable at about 0.3~mag d$^{-1}$. In the case of the long outbursts, the  rate varies between 0.3~mag~d$^{-1}$ and 0.5~mag~d$^{-1}$.
The average time in low brightness state at $V=17.0\pm0.2$ mag is 
about 1--5 days.

\begin{figure}
\includegraphics{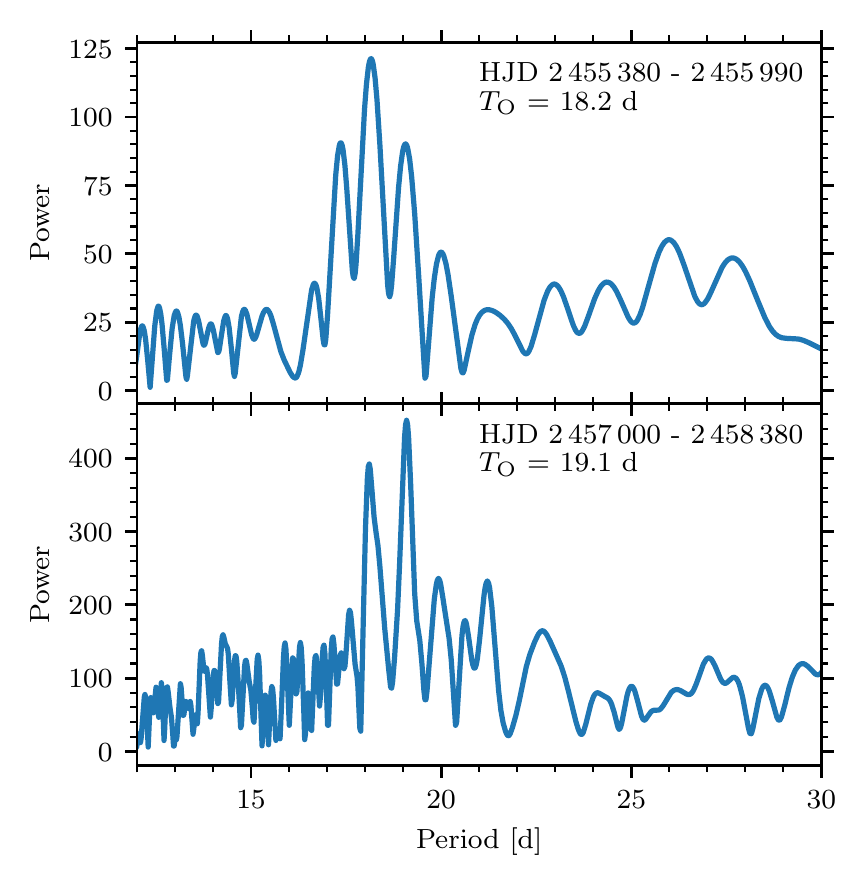}

\caption{Power spectrum of observations obtained on HJD 2\,455\,380--2\,455\,990 based on AAVSO data (\textit{top panel}) and power spectrum of observations obtained on  HJD 2\,457\,000--2\,458\,380 based on AAVSO, ASAS-SN and ZTF data (\textit{bottom panel}).
}
\label{F:PWRS}
\end{figure}

The outbursts' occurrence during the active phase is quasi-periodic. The power spectrum of AAVSO observations between HJD 2\,455\,380--2\,455\,990 (2010 July 2 - 2012 March 3) gives outburst recurrent time of $T_\mathrm{O} = 18.2(4) \: \mathrm{d}$ and the power spectrum of AAVSO, ASAS-SN and ZTF observations between HJD 2\,457\,000--2\,458\,380 (2014 December 8 - 2018 September 18) gives  $T_\mathrm{O} = 19.1(2) \: \mathrm{d}$, both power spectra are shown in Fig.~\ref{F:PWRS}.

So far, there have been three standstills observed - the first one occurred between 
HJD 2\,456\,160--2\,456\,995 (2012 August 20 - 2014 December 3)
and lasted about 835 days. The second was between 
HJD 2\,457\,231--2\,457\,275 2015 July 27 - 2015 September 9
and lasted only $\sim$44 days,
and the most recent one started on HJD 2\,458\,402 (2018 November 10) and did not come to an end at the time of writing.
The most recent standstill is the longest one so far, with a length greater than $1500$ days.
There are, however, gaps in the observational data lasting $\sim 100$ days around the times of solar conjunctions (see Fig.~\ref{F:01}), and it is, therefore, possible that standstills could be interrupted during these gaps by a short active phase or that some short standstills were not detected.
The brightness of the system during standstill is $V \approx $ 15.4~mag, that is  $\sim$ 1.7~mag bellow the maximal brightness of outbursts. The standstill states are occasionally interrupted by the occurrence of outbursts or outburst/dip pairs. The maximal brightness during these outbursts is the same as of the ones occurring during active states, but they occur only sporadically.

\begin{table}[!tbp]
\caption{The outburst characteristics}
\label{Tab:Out}
\begin{center}
\begin{tabular}{cc|cc}
\hline
Parameters  & Unit     & Short &  Long        \\
\hline\noalign{\smallskip} 
$T_{\mathrm{d}}$ & d &   $11$ & $16-19$       \\
FWHM  &  d &  $6$  &  $10-14$       \\    
$A$     & mag &    $2.0$ & $2.5 $         \\
$\dot{m}_\mathrm{i}$ & mag d$^{-1}$ & $-0.5$ &   $-0.5$         \\
$\dot{m}_\mathrm{d}$ & mag d$^{-1}$ & $0.3$   & $0.3-0.5$          \\
$T_{\mathrm{max}}$ &  d & $0$   &   $6-9$      \\
\hline
\end{tabular}
\end{center}

\tablecomments{
Meaning of parameters: \\
$T_\mathrm{d}$ - duration,
$A$ - amplitude,
$\dot{m}_\mathrm{i}$ - increase rate,
$\dot{m}_\mathrm{d}$ - decrease rate,
$T_{\mathrm{max}}$~-~time duration at the maximum.
}
\end{table} 

\begin{figure*}
\includegraphics{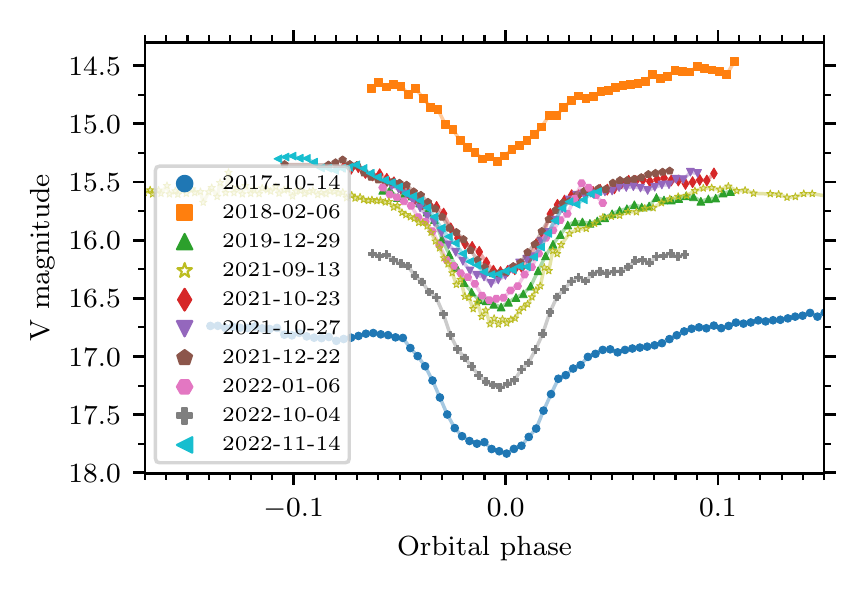}
\includegraphics{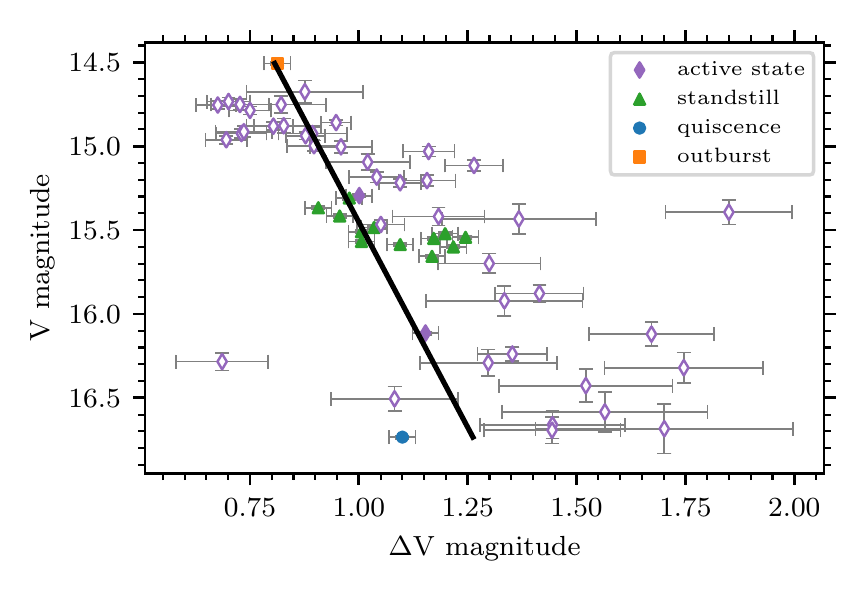}
\caption{{\it Left:} Eclipse light curves of AY~Psc obtained during 2017--2022 at \ond\ (filled-in symbols) and OAN SPM (empty stars). The color of light curves corresponds to the color of triangles in the top panel of Fig.~\ref{F:01}. The plot demonstrates how the system brightness and the eclipse depth and shape varied throughout our observational campaign. 
{\it Right:} The relation between the eclipse depth (horizontal axis) and out-of-eclipse brightness (vertical axis) in different activity states based on \ond{}, OAN SPM both marked by filled-in symbols), and  Exmouth (marked by empty symbols) observations. The linear fit of all points is marked by the black line. Fitting was performed using the orthogonal distance regression method to account for uncertainties in both $V$ and $\Delta \mathrm{V}$.
}
\label{fig:EclipseLC:Ond}
\end{figure*}

In Fig.~\ref{fig:EclipseLC:Ond} left, 
we collected eclipse profiles in different brightness states of the system obtained with low photometric errors ($\lesssim 0.03$ mag).
The C magnitudes  were transformed to the V-band using corresponding V-band magnitudes of several field stars and verified by comparison with nearly simultaneous AAVSO observations.
Fig.~\ref{fig:EclipseLC:Ond} right shows a relation between the out-of-eclipse system brightness and the eclipse depth for the active state and  standstill. 
To accommodate for out-of-eclipse changes in brightness and for different phase coverage of the eclipse light curves, only the phase interval $\varphi \in$~[$-$0.1; 0.1] was used to determine the out-of-eclipse brightness and the eclipse depth. The out-of-eclipse brightness was determined as the brightest measured magnitude from this phase interval, and the eclipse depth was determined as the difference between the faintest and the brightest magnitude.
During the active state, the eclipse depth gets smaller as the out-of-eclipse brightness increases (Fig.~\ref{fig:EclipseLC:Ond}, purple symbols) and shows a tendency for a linear dependence between them. The dependence is shown by a solid line described 
by relation 
\begin{equation}
    \Delta V = 0.20(4) \times V - 2.16(54),
\end{equation}
which was determined as the best linear fit of the presented data using the orthogonal distance regression method \citep{1990CM..112..189} which, unlike least square method, takes into account uncertainties of both quantities of the fitting function.
Similar behavior was also observed in some NL systems \citep{2020MNRAS.497.1475S}. 
The eclipse depth during standstill (the green symbols at Fig.~\ref{fig:EclipseLC:Ond}) is about $1.1$~mag, however, with a smaller dispersion of about $\sim0.2$~mag.

\subsection{Ephemeris}

\begin{figure}[t]
\includegraphics{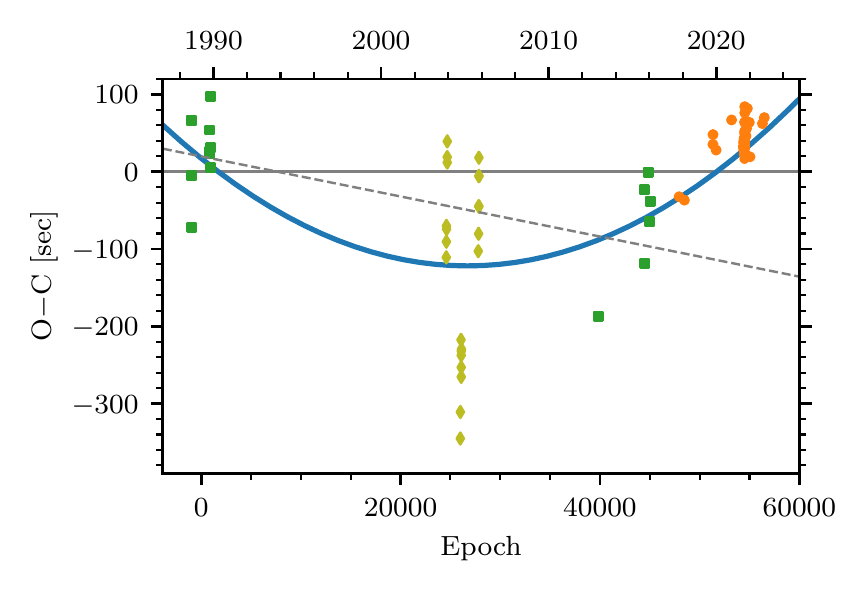}
\caption{The historical $O\!-\!C$ diagram for the mid-eclipse times of AY~Psc since its discovery. The diagram was constructed using the linear ephemeris given by Equation~\ref{EQ:EFL:NEW}, which is represented by the grey horizontal line in the plot.
The minima previously published by \cite{1990A&A...238..170D} and \cite{2017RAA....17...56H} are denoted by green squares, newly-derived mid-eclipse times from light curves of \cite{2009NewA...14..330G} are denoted by yellow diamonds, and our new minima are denoted by orange circles. 
The continuous blue curve corresponds to the best quadratic fit of the data described by Equation~\ref{EQ:EFQ:NEW}, and the grey dashed line represents the best linear fit derived by \cite{2017RAA....17...56H}.}
\label{ayoc}
\end{figure}

The first linear light elements of AY~Psc were derived by \cite{1990A&A...238..170D}:
\begin{equation}
    \mathrm{Pri.Min.} = \mathrm{HJD}\; 2\,447\,623.3463 + 0\fd2173209\times E,
\end{equation} 
where HJD $2\,447\,623.3463$ is the initial epoch in heliocentric Julian date, $0\fd2173209$ is the orbital period of the system and $E$ stands for the number of epochs.
The elements were later improved by \cite{2009NewA...14..330G}: 
\begin{equation}
    \mathrm{Pri.Min.} = \mathrm{HJD} \; 2\,452\,966.1728(4) + 0\fd217320(1)\times E.
\end{equation}
Finally, based on a four-year scale, \cite{2017RAA....17...56H}  derived the light elements as 
\begin{equation}
    \mathrm{Pri.Min.} = \mathrm{HJD} \; 2\,447\,623.34628(2) + 0\fd21732061(1)\times E.
\end{equation}
They also noted that the residuals from this ephemeris show some deviations 
and the orbital period does not seem to be constant. 
Several additional eclipse timings for AY~Psc have been reported in the literature. Besides those minima given in Table~\ref{Tab:ecl} and Table~\ref{Tab:ecl_TESS}, we used previous mid-eclipse times obtained by \cite{1990A&A...238..170D} and \cite{2017RAA....17...56H}.
A total of 155 CCD times were included in our period study, and we derived new linear ephemeris
\begin{equation}
    \label{EQ:EFL:NEW}
     \begin{array}{l}
    \mathrm{Pri.Min.} = \mathrm{BJD} \; 2\,447\,623.34668(13)\, + \\
  \hspace{16 mm}  + 0\fd217320641(3)\times E,
      \end{array}
\end{equation} 
where the initial epoch is given in barycentric Julian date.
The data, however, are better described by a quadratic ephemeris and, therefore
a quadratic ephemeris with a period increase was derived:  

\begin{equation}
    \label{EQ:EFQ:NEW}
  \begin{array}{l}
    \mathrm{Pri.Min.} = \mathrm{BJD} \; 2\,447\,623.34700(8) \, + \\ 
    \hspace{16 mm} + \, 0\fd217320523(8)\times E \, +  \\
    \hspace{16 mm}  +\, 2.3(1) \, 10^{-12}\times E^2,
  \end{array}
\end{equation} 
where $0\fd217320523$ is the orbital period at epoch $E=0$ and $2.3\,\times 10^{-12}$~d is the quadratic term describing the non-linear changes in the $O\!-\!C$ diagram. The period increase was also reported by \cite{2023MNRAS.519..352B}, who used eclipse timings listed by \cite{1990A&A...238..170D} and \cite{2017RAA....17...56H}, eclipse epoch given by \cite{2009NewA...14..330G} and TESS data to derive quadratic term of $3.69(2)\,\times 10^{-12}$~d.
The quadratic term in Equation \ref{EQ:EFQ:NEW}  indicates a slow continuous period increase 
with a rate of  $dP/dt = +7.6(5)\times10^{-9} \mathrm{\; d \; year}^{-1}$.
The complete $O\!-\!C$ diagram since the discovery of AY~Psc as a variable star is plotted in Fig.~\ref{ayoc}. 
Our new observations extended the time span by over 1800 days corresponding to 8300 epochs. 
 Excluding a part of data with a high scatter
around the epoch of $\sim25~000$ does not change our quadratic fit out of its 1$\sigma$ uncertainty.
It should be noted that determination of the center of the eclipse of the WD is difficult using only observations with a low time resolution, for which it is not possible to define the times of the eclipse ingress and egress. 
The observed minimum brightness does not exactly correspond to the mid-eclipse of the WD. Nevertheless, it is clearly visible that these times have not followed a simple linear ephemeris during the past 30~years.
Assuming a conservative mass transfer between components, the period increase, and stellar masses 
 determined below, one can compute an upper limit of a long-term averaged mass transfer rate from the secondary component to WD $\dot{M}$ using the equation 
 \begin{equation}
    \label{EQ:M_P_DOT}
    \dot{M} = \frac{\dot{P}}{3P} \, \frac{q \, M_1}{1-q},
\end{equation}where $M_1$ is the mass of the primary and $q$ is the mass ratio \citep[see][equation 9.5b]{Warner:1995aa}.
  Equation \ref{EQ:M_P_DOT} gives  $\dot{M} = 1.05(15) \times 10^{-8} \mathrm{\; M}_{\sun} \mathrm{\; year}^{-1}$.

\begin{figure}
   \centering
   \includegraphics{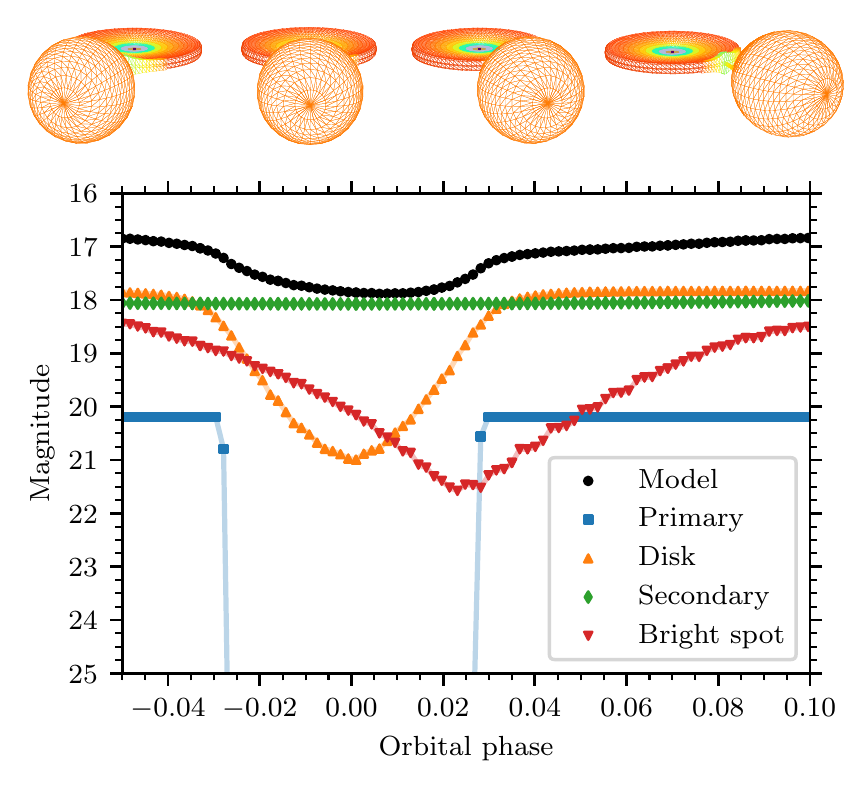}
   \caption{The contribution of different components to the total flux. The top side of the plot presents an artistic view of the system in corresponding orbital phases. The lack of continuity in the bright spot curve is due to a pixelation of our model. }
 \label{F:Model}
\end{figure}

\begin{figure*}[t]
 \setlength{\unitlength}{1mm}
\begin{center}
\includegraphics{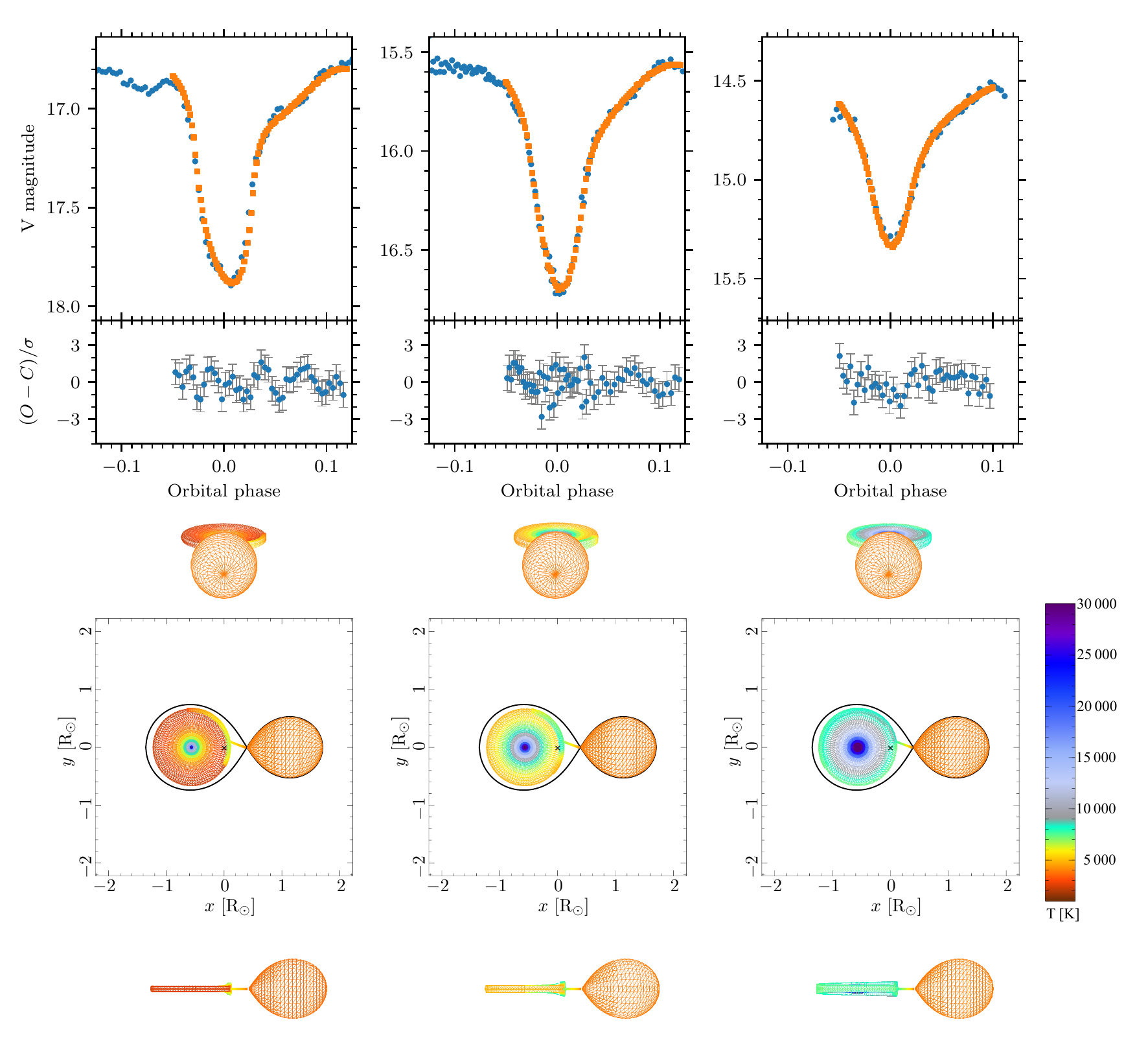}
\end{center}
\caption{{\it Upper part}: Results of the light curve fit of AY~Psc eclipses in quiescence/low ({\it left}), standstill  ({\it middle})
and in outburst/high ({\it right}) state. 
Mid-eclipse times of the displayed light curves are  BJD~2458041.48048, BJD~2459471.01677, and BJD~2458156.22572, respectively.
The blue points show the observational data presented in Fig.~\ref{fig:EclipseLC:Ond}, left. The orange squares mark the modeling light curves corresponding to the best fits. The \ocs\ residuals for the fits are given. Note the different
values on the vertical axis.
{\it Bottom}: Models of AY~Psc in quiescence ({\it left}), standstill ({\it middle}) and in outburst ({\it right}).  The colors mark
the effective temperature in Kelvin of the system elements ({\it right bar}). The upper panels correspond to orbital phase $\varphi=0$, the middle ones are views on the geometry of the system at the orbital phase of $\varphi=0.25$ from 
the inclination angle $i=0.0$, and the lower panels correspond o $\varphi=0.25$  and inclination angle $i=90.0$ degree.  
} 
\label{F:MCF1}
\end{figure*}

\begin{table*}[!tbp]
\caption{System parameters from modeling of light curves of eclipses.}
\label{tab:BestPar}
\begin{center}
\begin{tabular}{lllccc}
\hline
 &  &  &    \multicolumn{3}{c}{\bf System brightness state}  \\
\multicolumn{3}{l}{\bf Adopted parameters} & low$^\ast$  &  high$^\star$ & standstill$^\diamond$          \\   
\hline
\multicolumn{3}{l}{$P_{\mathrm{orb}}$}  &  \multicolumn{3}{c} {18\,776.49 s}   \\ 
\multicolumn{3}{l}{$E(B-V)$ }           &  \multicolumn{3}{c} {0.05}    \\
\multicolumn{3}{l}{Distance}            &  \multicolumn{3}{c} {730 pc}  \\    
\hline\noalign{\smallskip}
\multicolumn{4}{l}{{\bf Variable and their best values}} \\  \hline\noalign{\smallskip}

\multicolumn{3}{l}{ $i$ [degree]}                 & { 74\fdg8(7) } & fixed & fixed  \\
\multicolumn{3}{l}{ $M_{\mathrm{WD}}$ [\ms]}      & {0.90(4) } & fixed & fixed  \\
\multicolumn{3}{l}{ ${T}_{\mathrm{WD}}$ [K] }     & {27\,600(12\,000)  } & fixed & fixed\\ 
\multicolumn{3}{l}{ $T_{2}$ [K] }                 &  4100(50) & fixed  & fixed  \\
\multicolumn{3}{l}{ $q  $}                        &  0.50(3) & fixed &  fixed \\
\noalign{\smallskip}\hline\noalign{\smallskip}
\multicolumn{3}{l}{ $\dot{M}$ [$10^{-10}$ \ms\ year$^{-1}$] (model) } & 2.4(1) &  96.8 &  9.1 \\

\hline\noalign{\smallskip}
\multicolumn{4}{l}{{\bf Parameters of the disk}} \\ 
\hline
\multicolumn{3}{l}{ $R_{\mathrm{d, in}} \equiv R_{\mathrm{WD}}$ [\rs] }  &  0.009  & fixed & fixed\\
\multicolumn{3}{l}{ $R_{\mathrm{d, out}}$ [\rs]} & 0.67(1)    & 0.67  & 0.67  \\
\multicolumn{3}{l}{ $z_{\mathrm{d, out}}$ [\rs]} &  0.047(5)  &  0.049  & 0.061  \\  
\multicolumn{3}{l}{ $\gamma_{\mathrm{disk}}$ }   &  1.223(0.500)     &  0.00 & 0.48  \\  
\multicolumn{3}{l}{ EXP }                        &  0.250(1)     &  0.238 & 0.228   \\ 
\noalign{\smallskip}\hline\noalign{\smallskip}
\multicolumn{3}{l}{{\bf The bright spot/line}}&  & &  \\ \hline\noalign{\smallskip}
\multicolumn{3}{l}{length spot ($\varphi_{\mathrm{min}}+ \varphi_{\mathrm{max}}$) }  &\multicolumn{1}{c} {128\fdg9(9.5)} &  \multicolumn{1}{c} {139\fdg3} & \multicolumn{1}{c}{129\fdg4} \\
\multicolumn{3}{l}{width spot [\%] }  &\multicolumn{1}{c} {$9.5_{-8}^{+5}$} & \multicolumn{1}{c} {22.9}&  \multicolumn{1}{c} {13.2}   \\
\multicolumn{3}{l}{Z excess spot ($z_{\mathrm{s, max}}/z_{\mathrm{d,out}}$)  }  &\multicolumn{1}{c} {1.27(0.55)} & \multicolumn{1}{c} {1.65} & \multicolumn{1}{c} {1.01}  \\
\multicolumn{3}{l}{Spot Z slope }  &\multicolumn{1}{c} {3.9(1.5)}  & \multicolumn{1}{c} {1.4} &\multicolumn{1}{c} {2.5}  \\
\multicolumn{3}{l}{Shift  spot ($\varphi_{\mathrm{min}}$)}  &\multicolumn{1}{c} {-30\fdg3$(3.0)$} & \multicolumn{1}{c} {-20\fdg7} &\multicolumn{1}{c} {-42\fdg5}   \\
\multicolumn{3}{l}{Shift $T_{\mathrm{s, max}}$ ($\varphi(T_{\mathrm{s, max}})$)}  &\multicolumn{1}{c} {-6.8\fdg0(2.0)} & \multicolumn{1}{c} {-0\fdg2} &\multicolumn{1}{c} {-11\fdg3}    \\
\multicolumn{3}{l}{Temp. excess spot ($T_{\mathrm{s, max}}/T_{\mathrm{d,out}}$)  }  &\multicolumn{1}{c} {2.7(3)} & \multicolumn{1}{c} {1.2}  &\multicolumn{1}{c} {1.8} \\
\multicolumn{3}{l}{Spot Temp. slope (left) }   &\multicolumn{1}{c} {$0.18_{-0.07}^{+0.15}$}  &\multicolumn{1}{c} {1.32} &\multicolumn{1}{c} {1.0}    \\
\multicolumn{3}{l}{Spot Temp. slope (right) }  &\multicolumn{1}{c} {$0.41_{-0.1}^{+0.2}$}& \multicolumn{1}{c} {0.25}  &\multicolumn{1}{c} {0.4}   \\

\hline\noalign{\smallskip}
\multicolumn{3}{l}{{ $\chi^2$/dof}}  &\multicolumn{1}{c} {0.74/31} & \multicolumn{1}{c} {0.79/29} &\multicolumn{1}{c} {0.94/47}  \\ 
\hline\noalign{\smallskip}
\multicolumn{4}{l}{{\bf Calculated parameters}} \\ 
\hline\noalign{\smallskip}
\multicolumn{3}{l}{ $\dot{M}$ [$10^{-10}$ \ms\ year$^{-1}$] (eq.\ref{Lum})} & 2.4  & 135.7 & 15.8 \\ 
\hline\noalign{\smallskip}
\multicolumn{3}{l}{$a$ [\rs]}  &\multicolumn{3}{c}{1.68 }  \\
\multicolumn{3}{l}{$M_{2}$ [\ms] }  & \multicolumn{3}{c}{0.45 } \\
\multicolumn{3}{l}{$R_{2}$ [\rs] }  & \multicolumn{3}{c}{0.54  }\\

\hline

\end{tabular}
\end{center}

\tablecomments{
 Numbers in brackets for the variables are uncertainties defined as 1$\sigma$ of the Gaussian function approximation used to describe the 1-dimension $\chi^2$ function.
 The {\sc Gaia} distance uncertainties are less than 3\%. They extend only errors of the mass transfer rate and/or the temperature of the secondary by a similar range of values. Other parameters of the model vary only inside their errors.
  Mid-eclipse time: $^\ast$~-~BJD 2458 041.48048; $^\star$ - BJD 2458156.22572;  $^\diamond$ - BJD 2459471.01677 
  }

\end{table*}

\subsection{Eclipse light-curve modeling and system parameters}
\label{sec:LCfit}

The AY~Psc system shows strong activity, and the time in quiescence is significantly shorter than the length of outbursts or standstills (see Section~\ref{S:LTV}).  The quiescence lasts in a range of about 5--23 orbital periods. The full time of outbursts and the time 
in the plateau phase of outbursts lasts on average up to $\sim$80 and $\sim$40 orbital periods, respectively.
In contrast to the active state of AY~Psc, the standstill duration is
very long when compared with the orbital period of the systems (months or even years).
Generally, the disk in the system is not in a steady state
if we consider its evolution along  whole time intervals of quiescence or during outbursts.  
It is more likely to resemble the steady state during a standstill.
Such fast activity of the system complicates the analyses of the physical properties 
of the accretion disk significantly. Nevertheless, the eclipse light curve profile 
(ELCP) in different states of the system allows us to probe intensity or effective 
temperature distribution along the disk radius and to try to determine the system parameters. 
To do this, we attempt to reproduce the collected photometric observations of the system 
eclipse light curves in quiescence, standstill, and during an outburst using the tool 
developed by \citet{2013A&A...549A..77Z}.  The detailed
description of the model can be found in our recent papers of \citet{2020MNRAS.497.1475S} and \citet{2021A&A...652A..49K}. 
Here we note only that the light curve is a superposition of flux contribution of four main components: the white dwarf primary, 
the secondary star, and the accretion disk with the bright spot formed at the disk/stream impact region.
The depth and shape of the eclipse depend on the flux from the secondary and the brightness distributions on the disk and
the bright spot. The bright spot affects more the descending branch of ELCP, i.e., before reaching the minimum. The side of ELCP
after the minimum is formed mainly by contribution from the disk and partly by the underlying secondary.
Only after the orbital phase of $\varphi \approx0.05$ does the bright spot affect the total flux of ELCP significantly.  Fig.~\ref{F:Model}
shows the contributions of different components to the resulting eclipse light curve. Below, we give only  the specifications of the model
connected with the considered case.

\begin{figure*}
 \setlength{\unitlength}{1mm}
\begin{center}
\begin{picture}(150,177)(0,0)
\put (-15,0) {\includegraphics[width=18.2cm,bb = 20 50 980 980, clip=]{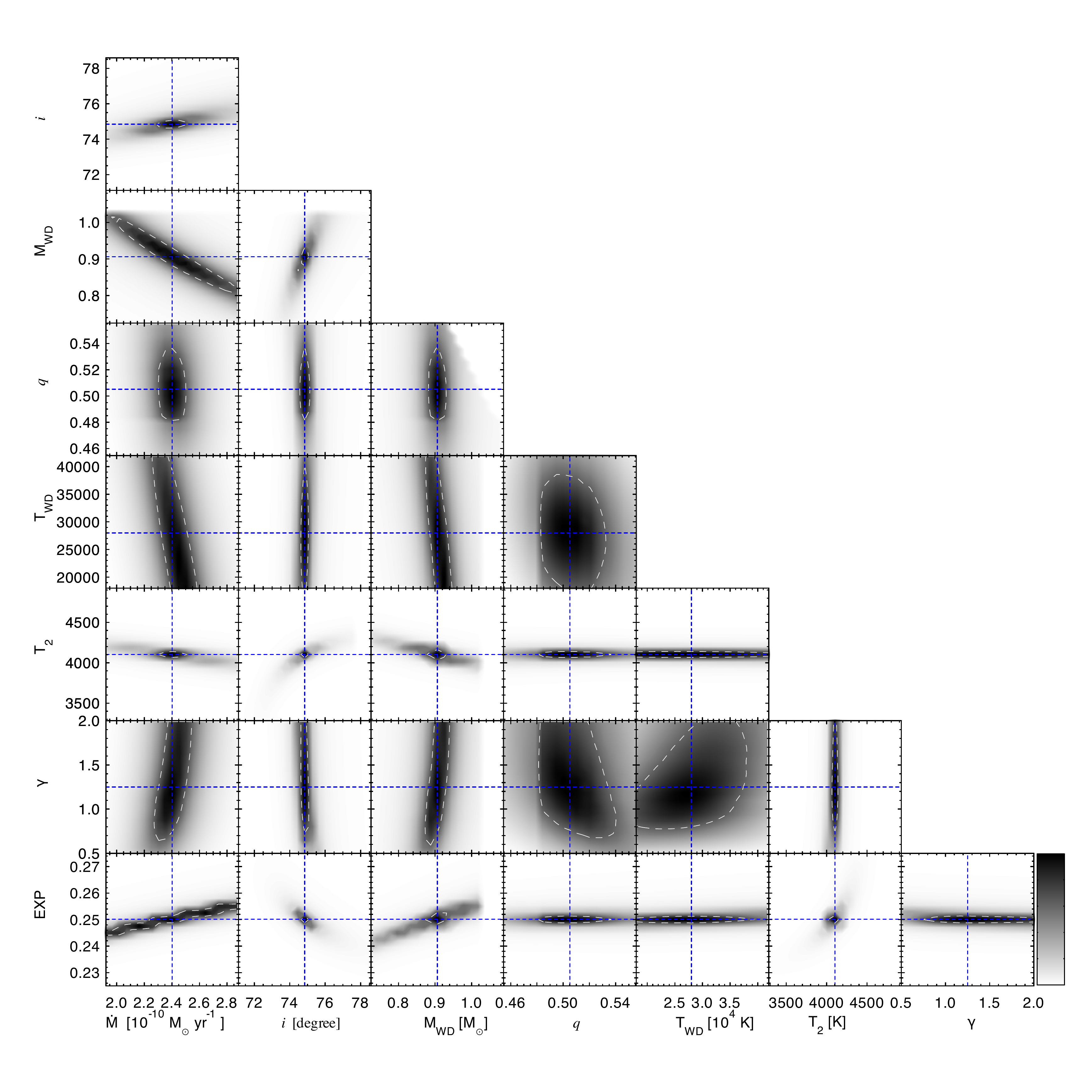}}
\put (167,30) {1}
\put (167,7) {0}
\end{picture}
\end{center}
\caption{Posterior distributions of some parameters (see Table~\ref{tab:BestPar}) of the fit for the AY~Psc light curve in the low brightness state. The grey scale shows normalized $\chi^{-2}$.
The blue dashed lines mark the best fit values of parameters. 
The white long-dashed lines correspond to 1$\sigma$ errors of parameters.
}
\label{F:Errors}
\end{figure*}

\begin{figure}
\begin{center}
\includegraphics{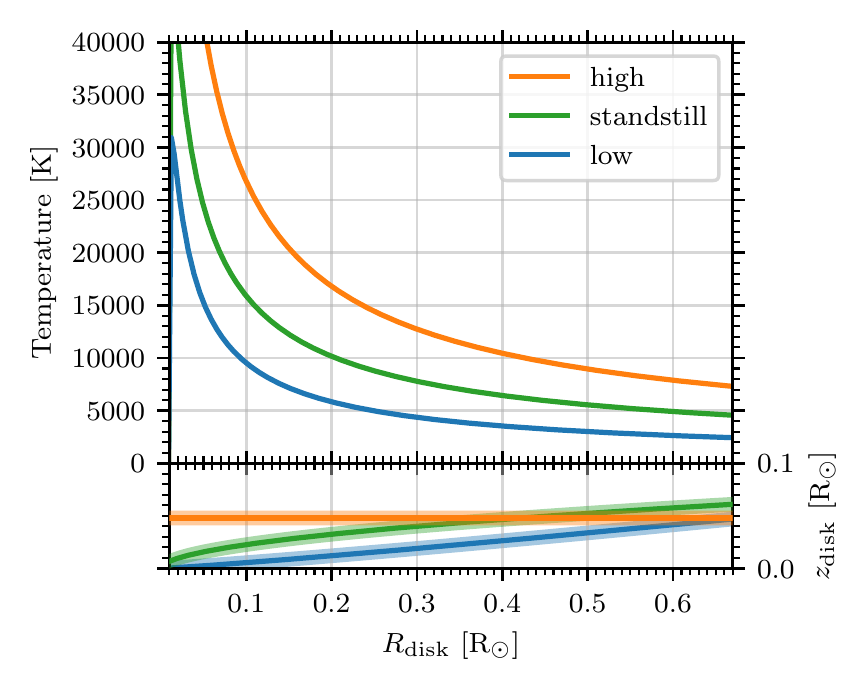}
\end{center}
\caption{  Temperature (top panel) and height of accretion disk (bottom panel) from the model fitting. 
The 1$\sigma$ uncertainties of the disk height are marked in the bottom panel by light-shaded areas.
The orange, green, and blue lines correspond to the high, standstill, and low state of the system.}
\label{F:TZModel}
\end{figure}

 Foremost, we take into account that the disk contribution in the total flux from the system is lower in quiescence than in
outbursts or standstills. The flux from the far side of the secondary is always the same, but its contribution to the total flux is
significantly larger in quiescence. Therefore, we used the eclipse light curve in the low-brightness state of the system obtained on October 14, 2017 (the eclipse time $\mathrm{BJD}\;2\,458\,041.48048$, the last light-curve point was obtained at $\mathrm{HJD}\;2\,458\,041.51871$) to find the system parameters. We verified our calibration of the flux in the $V$-band using AAVSO data obtained close to our observations at $\mathrm{JD}\;2\,458\,041.59532$ where $V=16.73(11)$ mag was measured about two hours after the used eclipse.  We adopted the published values of the orbital period, the distance, and the interstellar extinction as the fixed parameters. All other parameters of the model, described below and presented in Table~\ref{tab:BestPar},  were free.  
 
 We assume that the disk radiates as a black body at the local effective temperature with radial distribution across the disk given by the equation:
\begin{equation}
 \begin{array}{lll}
T_{\rm eff}(r) &  = &T_0 \left\{ \left(\frac{r}{R_{\rm WD}}\right)^{-3} \, \left(1-
\left[\frac{R_{\rm WD}}{r}\right]^{1/2}\right)\right\}^{EXP},  \\
T_0 & = & \left[\frac{3GM_{\rm WD}\dot{M}}{8\pi\sigma R_{\rm WD}^3}\right]^{1/4},
\end{array}
\label{Tempeq}
\end{equation}

\noindent
where $M_{\mathrm{WD}}$ and $R_{\mathrm{WD}}$ are the mass and the radius of the WD, respectively, 
$\dot{M}$ is the mass-transfer rate, 
$G$ is the gravitational constant and 
$\sigma$ is the Stefan--Boltzmann constant. 
 In the standard accretion disk model the radial temperature gradient $EXP$ = 0.25 \citep[equation 2.35]{Warner:1995aa},
but we allow it to slightly deviate from this value, similarly to how it was done  in \citet{2010ApJ...719..271L}.   Generally, this departure from the standard temperature distribution can be presented as the multiplication of the standard steady-state model 
$T_{\rm eff, SSD}(r)\propto r^{-3/4}$ to a slowly changing function $f(r)\propto r^{\delta}$ reflecting the order of a deviation  from the steady-state disk.
The  mass transfer rate
corresponding to the luminosity state of the disk  can be obtained from the relation
\begin{equation}
L_{\rm disk} = \frac{1}{2}\frac{GM_{\rm WD}\dot{M}}{R_{\rm WD} },
\label{Lum}
\end{equation}
where $L_{\rm disk}$ is the total disk luminosity calculated  as
\begin{equation}
L_{\rm disk} = \sum_{i}^{N} S_{i}\sigma T_{\rm eff, i}^4,
\label{Lum1}
\end{equation}
where $N$ is the number of elements in the model of the accretion disk, $S_{i}$ and $T_{\rm eff, i}$ are an area and an effective 
temperature of an $i$-element, respectively.
 The disk limb-darkening also includes the following Eddington approximation
\citep{1980MNRAS.193..793M, 1980AcA....30..127P}.
We also used the complex model for the bright spot described in detail in \citet[see fig.~9 therein]{2021A&A...652A..49K} where 
all definitions of parameters can be found. Clearly, the true bright spot configuration is more complicated, and our models only give general impressions about brightness distribution in the stream/disk impact region. 
The free parameters of this fit were:   
the mass of the primary ($M_{\mathrm{WD}}$), 
the mass ratio ($q=M_2/M_{\mathrm{WD}}$), 
the mass transfer rate ($\dot{M}$),  
the system inclination ($i$), 
the outer radius of the accretion disk ($R_{\mathrm{d, out}}$),
the disk thickness at the outer radius ($z_{\mathrm{d, out}}$),  
the effective temperature of the secondary ($T_2$), and 
the accretion disk with a thickness 
\begin{equation}
z_\mathrm{d}(r) = z_\mathrm{d}(r_{\mathrm{out}})(r/r_{\mathrm{out}})^{\gamma_{\mathrm{disk}}},
\end{equation}
 where $\gamma_{\mathrm{disk}}$ is a free parameter for which we used the standard value of $\gamma_{\mathrm{disk}}= 9/8$ \citep[equation 2.51b]{Warner:1995aa} as the initial value. We fit only the orbital phase range 
of $\varphi \in [-0.05;+0.12]$ which is less affected by out-of-eclipse flickering of the disk, as  can be seen on  Fig.~\ref{fig:EclipseLC:SPM}. 

 As the first approach, we found by hand an approximate solution of the light curves with steady-state parameters of the disk. 
The mass transfer rate, in this case, can be considered as an average in the disk corresponding to the current luminosity state of the system.
After that, we searched for the minimum of the $\chi^2$ function with free  parameters mentioned above (see also Table~\ref{tab:BestPar}) 
using the gradient-descent method.
The values of the best-fit model are presented in Table~\ref{tab:BestPar}. The error of each fitted parameter  was calculated with the Gaussian approximation of the  $\chi^2$ function when other parameters were set at the best values. 
The result of the fit and
the observed-minus-calculated \ocs\ residuals are shown in Fig.~\ref{F:MCF1}, left panels. 
  As an example, in  Fig.~\ref{F:Errors}, we show two-dimensional error maps for some parameters.

 We found that the selected eclipsing light curve in quiescence was reproduced  by the accretion disk with parameters close to steady-state disk ($EXP$ = 0.250 and $\gamma_{\mathrm{disk}}\simeq 9/8$, see Fig.~\ref{F:Errors}, two bottom lines).  
The expected flux from the secondary is about $V=18.2$ mag, which is slightly lower than the magnitudes at the eclipse minimum (see Fig.~\ref{fig:EclipseLC:Ond}) in the quiescence.  
As we noted before, the main difference between ELCP in different system states is the depth of the eclipse.
Applying our light curve modeling technique to the high~\footnote{ the eclipse time $\mathrm{BJD} \; 2\,458\,156.22572$} and standstill~\footnote{ the eclipse time $\mathrm{BJD} \; 2\,459\,471.01677$} states (Fig.~\ref{F:MCF1}, middle and right panels)  with the system parameters obtained for the low state, we found a departure of the accretion disk from the steady-state:  The light curves in high brightness states were reproduced by the models with effective temperatures which are different from steady-state distribution and the height along disk radius is flatter (see Fig.~\ref{F:TZModel}). 

The mass transfer rate corresponding to the luminosity of the source changes drastically between the low and the high states at about two orders of magnitude.
The mass transfer rate in high state $\dot{M} \approx 1.4 \times 10^{-8} \mathrm{\; M}_{\sun} \mathrm{\; year}^{-1}$ is larger than the one derived from the orbital period changes, which is  $\dot{M} \approx 1.1 \times 10^{-8} \mathrm{\; M}_{\sun} \mathrm{\; year}^{-1}$. As the former corresponds to a mass accretion rate during a short event, namely the maximum of an outburst, and the latter corresponds to a long-term averaged value, this discrepancy can be expected. 

The mass transfer rate in standstill is  $\approx 2\times10^{-9}$~M$_\sun$~year$^{-1}$. This mass transfer rate is  slightly below   the critical mass transfer rate above which a cataclysmic variable  is hot and stable 
\citep[$\approx 7\times10^{-9}$~M$_\sun$~year$^{-1}$;][]{2018A&A...617A..26D}, however, it is in the range of  $\dot{M}$ estimations for Z Cam systems \citep[see fig.~3 therein]{2018A&A...617A..26D}.
The model uncertainties  of the fits in the high brightness state and standstill have the same order of values as in quiescence.

\section{Spectroscopy and Doppler tomography}
\label{Dopmaps}

\subsection{AY~Psc spectra}

\begin{table}[!tbp]
\caption{Radial velocity solution}
\label{T:VC}
\centering
\begin{tabular}{llll}
\hline
Method      &  $\upsilon_{\rm cor}$ [$\mathrm{km \, s}^{-1}$]  & $K$  [$\mathrm{km \, s}^{-1}$] & $\varphi_0$ \\
\hline
H$\alpha$, this paper   & $34.8(8.0)$ & $147.0(10.5)$  & $0.12(1)$ \\
H$\beta$, \citetalias{1993ApJ...403..743S} & $67(17)$ & $131(20)$  & $0.05(3)$ \\

\hline 
\end{tabular} 
\end{table}

\begin{figure}
   \centering
   \includegraphics{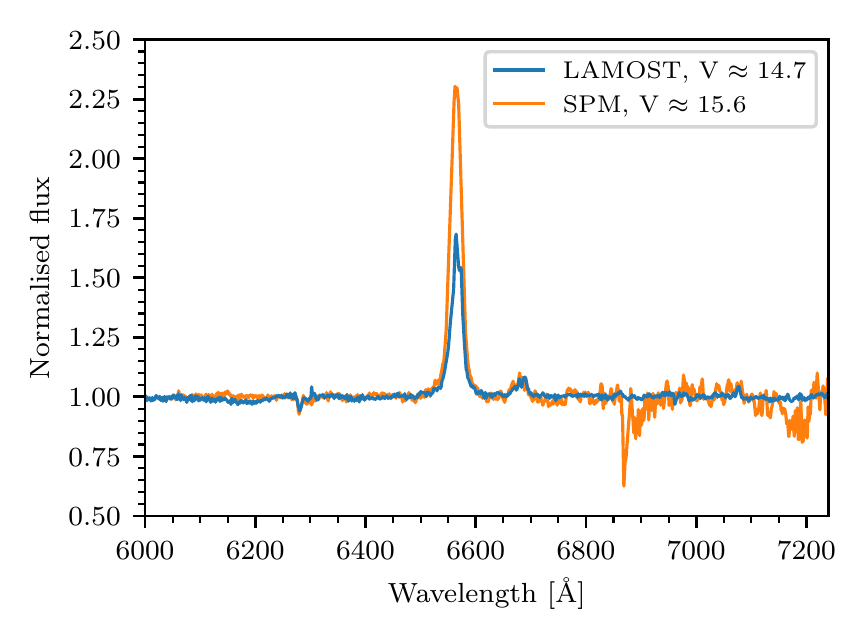}
   \caption{Normalised average spectrum from OAN SPM and normalized LAMOST spectrum.
  The wavelength range corresponds to the spectral range of OAN  SPM data. }
 \label{F:SP_N_C}
\end{figure}

\begin{figure}
   \centering
   \includegraphics{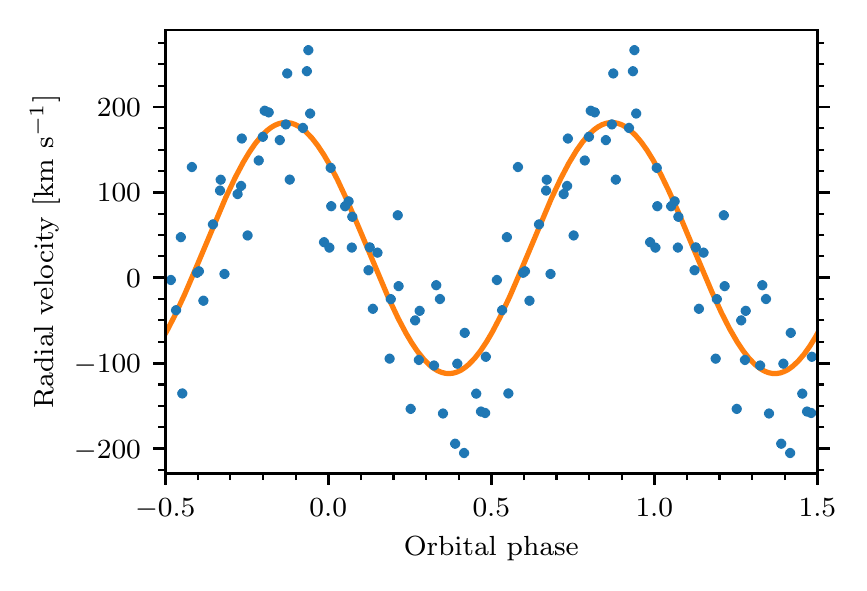}
         \caption{H$\alpha$ radial velocities found using the diagnostic diagrams. 
      The orange line marks the best fit of the data.
              }
         \label{F:02}
   \end{figure}

\begin{figure*}[t]
 \setlength{\unitlength}{1mm}
\centering
\includegraphics[width=16.9cm]{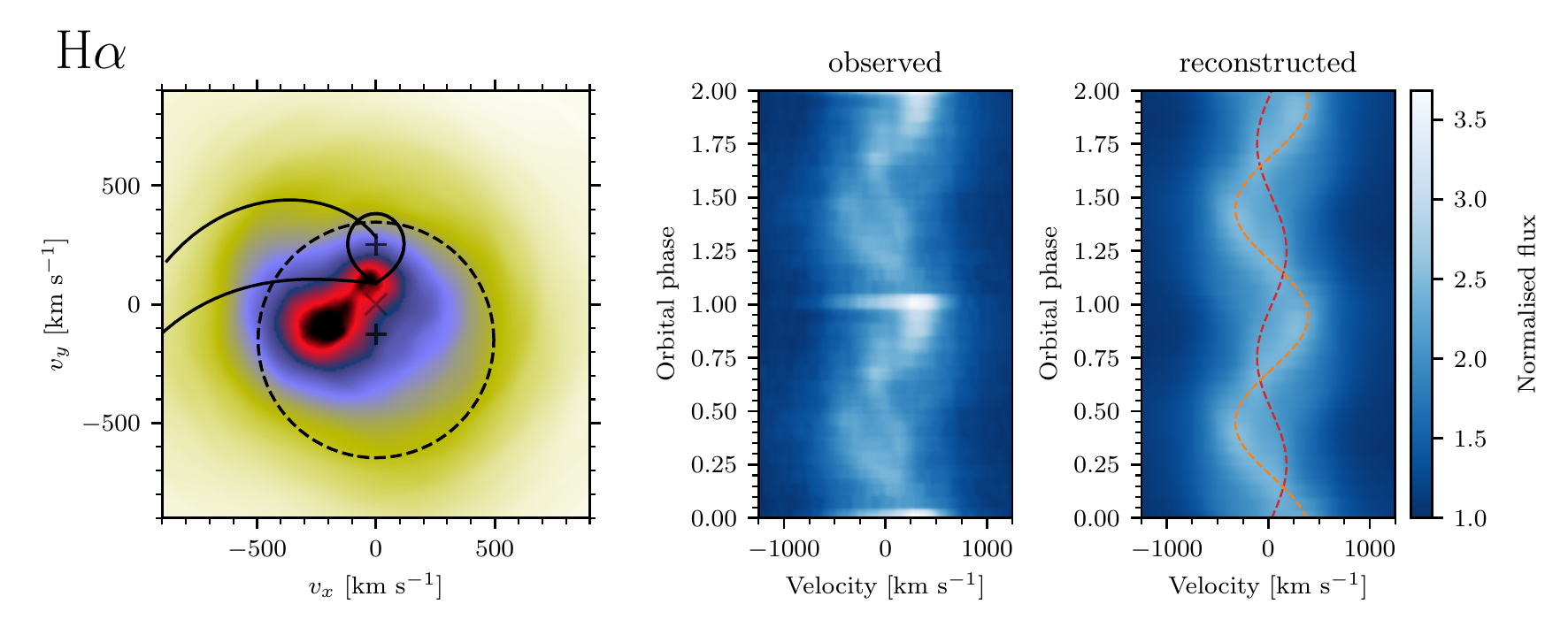}
\includegraphics[width=16.9cm]{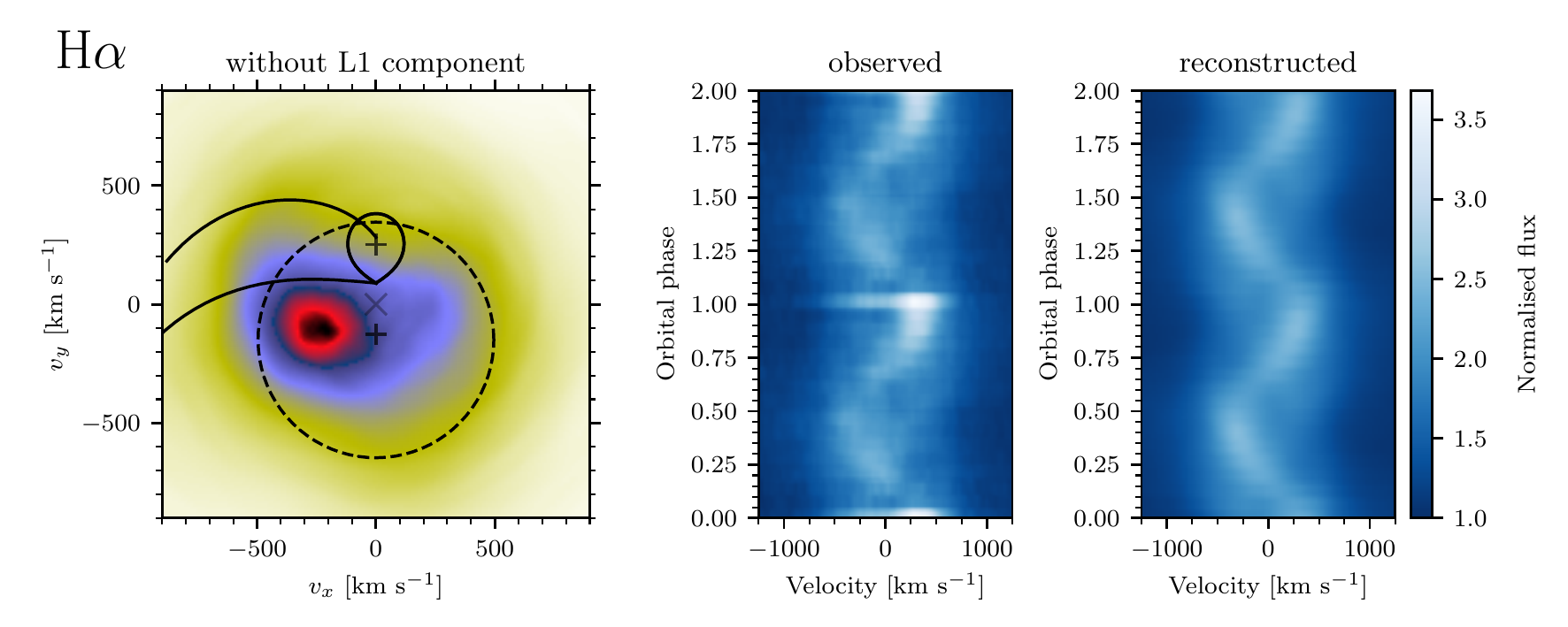}
\includegraphics[width=16.9cm]{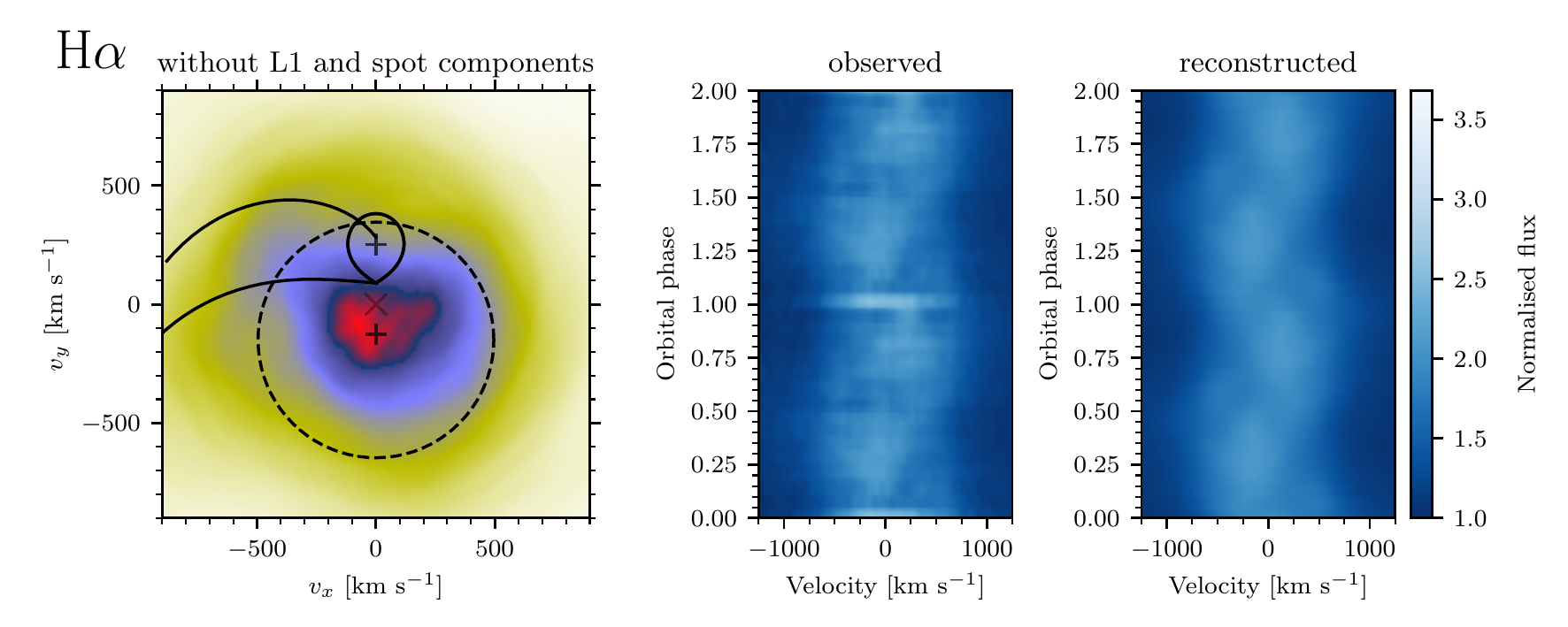}
\caption{Doppler maps for H$\alpha$
   line ({\it left panels}), and observed and reconstructed trailed spectra ({\it centre and right panels, respectively}).  The top left panel shows the Doppler map constructed using the  raw data. The middle left and the bottom left panels present the Doppler maps in which we removed the L1 and L1 + spot components, respectively (see description in the text). 
   The color of the Doppler maps corresponds to arbitrary units of emission intensity  (the yellow-blue-red-black palette corresponds to a change from a low to high intensity). The dashed circle marks the tidal limitation radius of the accretion disk. The red and orange dashed lines in the reconstructed trailed spectra of raw H$\alpha$ line ({\it top right panel}) mark the radial velocities of the `L1' and `spot' components, respectively. }
 \label{F:04}%
\end{figure*}

\begin{figure}[t]
   \centering
   \includegraphics{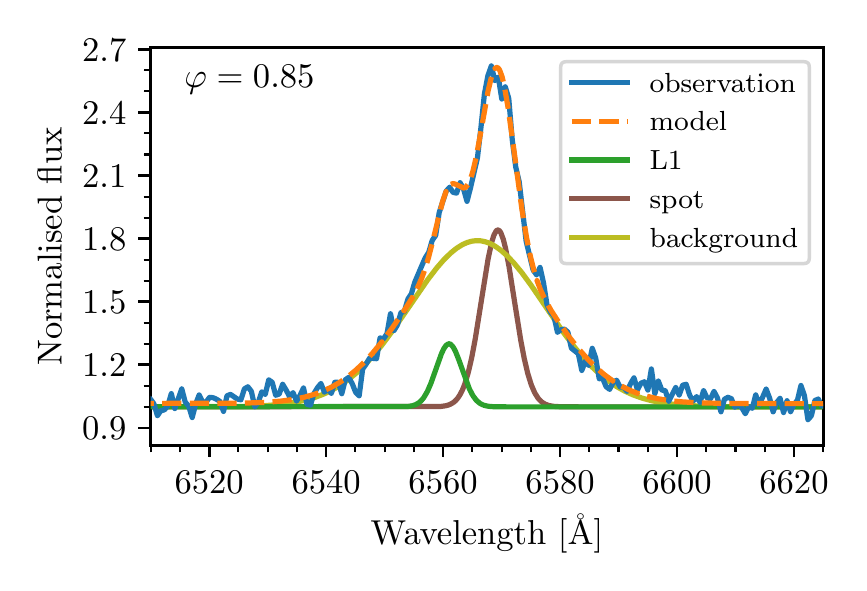}
         \caption{ Example of a three-Gaussian fits of the H$\alpha$ emission line profile  at the orbital phase $\varphi =  0.85$. The blue curve corresponds to the observed line profile. The green, brown, and yellow lines represent `L1', `spot', and background components (see the text), respectively. 
         The dashed orange line is the sum of those three components. }
         \label{F:3Gfit}
   \end{figure}

\begin{figure*}[t]
   \centering
   \includegraphics{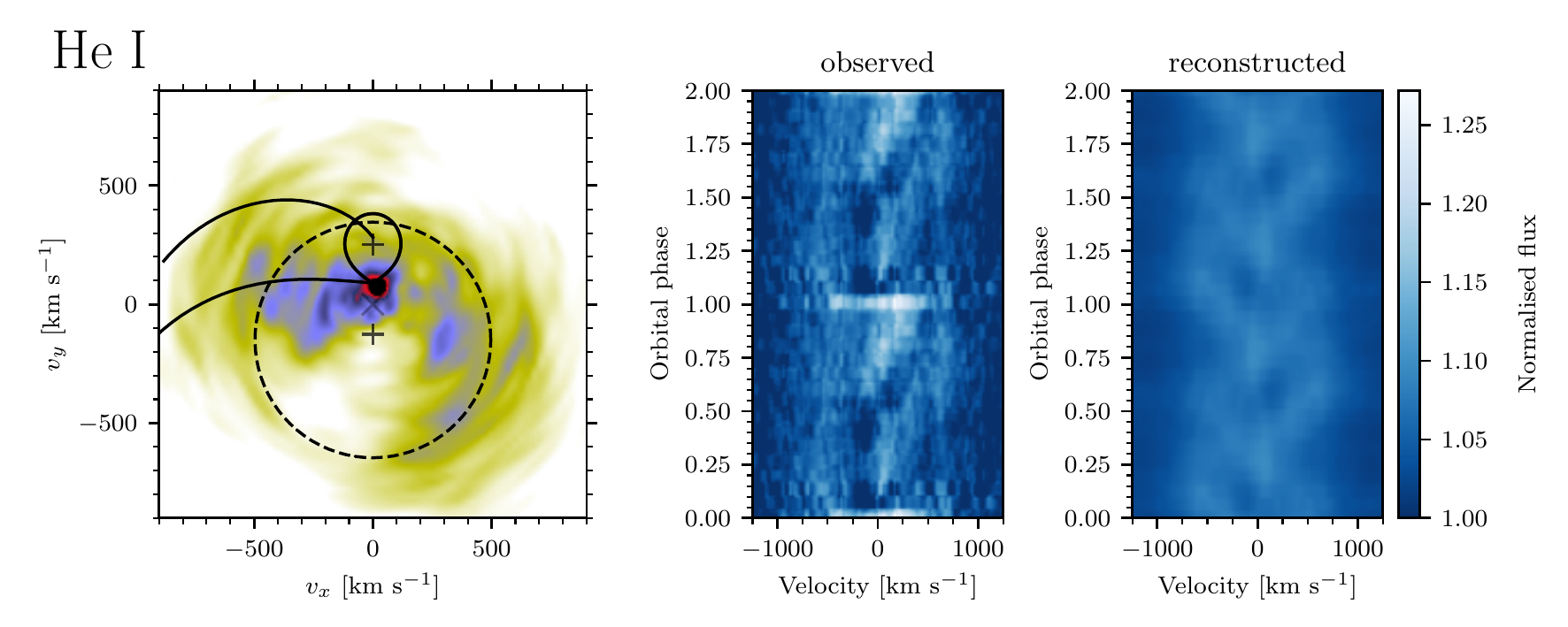}
   \caption{Doppler map for \ion{He}{1} line ({\it left panel}), observed and reconstructed trailed spectra ({\it right two panels}). The color of the Doppler maps corresponds to arbitrary units of emission intensity, the dashed circle marks the tidal limitation radius of the accretion disk. The corresponding orbital phase is marked.}
              \label{F:04b}%
\end{figure*}

Our time-resolved spectroscopy was obtained around  the H$\alpha$ line when the source was in the standstill state.
In the observed spectral region, there are strong H$\alpha$ and relatively weak \ion{He}{1}~6678 \AA\ emission lines. 
The phase-averaged  normalized spectrum from OAN SPM 
is shown in Fig.~\ref{F:SP_N_C} together with 
the same spectral region of the LAMOST spectrum for comparison. We covered
about four full orbital periods with uniform spectra distributed along the orbital phase. The phase-averaged H$\alpha$ 
line profile looks practically symmetric. Its FWZI and EW are 105.5 \AA\ and 35.2 \AA, respectively.
We also measured radial velocities of H$\alpha$ line using the double-Gaussian method described by  
\cite{1980ApJ...238..946S} and \cite{1983ApJ...267..222S}. We used Gaussians with the width of $1.5$ \AA\ and 
separated by $19.5$ \AA.  The resulting radial velocity curve 
is shown in Fig.~\ref{F:02}. We fitted the radial velocity curve with a model curve described by formula\begin{equation}
\label{EQ:12}
 \upsilon  =   \upsilon_{\rm cor} - K \sin\,[2\pi(\varphi - \varphi_0)], 
\end{equation} where $\upsilon_{\rm cor}$ is the systemic velocity, 
$K$ is the semi-amplitude of the radial velocity curve, 
and $\varphi$ is the orbital phase  calculated relative to the epoch 
$T_0$ = HJD 2\;459\;470.79867 which corresponds to the conjunction of the red and white dwarfs ($\varphi_0$=0),
when the L1 point velocity changes from negative to positive values.
The best fit is shown in Fig.~\ref{F:02}, and its parameters are listed in Table~\ref{T:VC}.
We found a good agreement of our results with those reported by \citetalias{1993ApJ...403..743S},
which are given in Table~\ref{T:VC}. Nevertheless,  detailed analyses of  variations of  the H$\alpha$ line profile required 
us to apply a multi-components model presented below.

\begin{table}[!tbp]
\caption{Multi-component model parameters}
\label{T:SCFb}
\centering
\begin{tabular}{l l c c }
\hline
 Parameter       & Unit      & \multicolumn{2}{c}{ Component}          \\
                 &           &  (1)         &     (2)            \\
                 &           &  `L1'        &  `spot'               \\
\hline
$\upsilon_{\rm cor}$ & km s$^{-1}$  &      $34.8$       & $34.8$          \\
$K$ &  km s$^{-1}$           &      $-145$     & $360$              \\
$\varphi_0$ &   --     &       0.0          & 0.196                  \\
$A$ & --         &      0.3        & 0.5     \\
$\sigma$ &  \AA      &       2.2      & 3     \\
\hline 
\end{tabular} 
\end{table}

\subsection{Doppler tomography of the  multi-component emission H$\alpha$~line }

We generated Doppler maps from our time-resolved spectra of the object using the code developed by \citet{1998astro.ph..6141S} which uses the maximum entropy method to compute the Doppler maps. The Roche lobe, position of the components, and stream were overplotted on the Doppler maps using a {\sc Python3} module {\sc PyDoppler} by \cite{2021ascl.soft06003H}.
The orbital phases ($\varphi \sim$ -0.05--0.05) around the eclipse minimum $\varphi_0 = 0$ were excluded 
from computations, and the zeroth orbital phase was well defined from the simultaneous photometric observations. 
The possible displacement between the zeroth orbital phase, when the secondary stays strictly between the 
primary and an observer, and a minimum of an eclipse does not exceed $\sim 0.005$ orbital phase, as follows 
from our light-curve modeling estimate. The system parameters from Section~\ref{sec:LCfit} are used to 
overlay  the secondary Roche lobe and position of the primary, the center of mass, and the ballistic and
Kepler stream trajectories. By circle, we show a tidally truncated radius ($R_{d, out, max} = 0.69 R_\odot$) of the disk which we estimated
using Equation~(2) from  \citet{2020A&A...642A.100N}.

The Doppler map of the H$\alpha$ line (Fig.~\ref{F:04}, top-left) and trailed spectra, based on the raw data,  clearly show the presence of at least two spectral components. One, located in the Doppler map at coordinates $V_{X}\approx$ 0 km s$^{-1}$, $V_{Y}\approx $ 100 km s$^{-1}$,  is bright, narrow and related to the Lagrangian point L1.  Another more extended component is located at $V_{X}\approx$ -300 km s$^{-1}$, $V_{Y}\approx $ -100 km s$^{-1}$, close to the standard bright spot position and named hereafter as the `spot' component. 
Both of them are located outside the accretion disk, whose outer radius is shown by the black dashed circle. 
They are placed above a very broad and heterogeneous background, the origin of which is discussed below.
In the trailed spectra, the `L1' components do not seem to vary with the orbital phase in its width and amplitude except for the time of the eclipse. Another component varies in its amplitude and width during the orbital phase and
has its maximal brightness just before the eclipse and a minimum at $\varphi \simeq 0.6$.

To determine the parameters of those spectral features, we fitted both those  components in such a way as to exclude their contribution from the trailed spectrum.
For each component, we used a simple Gaussian profile characterized by its peak flux $A$ and the standard deviation $\sigma$,
and the radial velocities $\upsilon$ dependent on the orbital phase:
 \begin{eqnarray}
   \mathrm{flux}(\lambda) & = & A \exp{\left[-\frac{1}{2}\left(\frac{\lambda-\lambda_0}{\sigma}\right)^2\right]},  \nonumber \\
   \lambda_0 & = & 6562.82 \; \left(1 + \frac{\upsilon}{c}\right) \, [\AA] ,
 \end{eqnarray}
 where $\upsilon$ is defined by Equation (\ref{EQ:12}). 
 As the first step, an eye estimation was applied to determine the parameters of the `L1' component. 
We accepted that the best fit means that the result of the component removal gives a similar flux from the position of the `L1'
component and its nearby vicinity. Afterwards, the parameters of the `spot' component at $V_{X}\approx$ -300 km s$^{-1}$,
$V_{Y}\approx $ -100 km~s$^{-1}$ in the Doppler map were determined by least-square fitting the spectra with a model consisting 
of three Gaussians representing both  components 
and the rest of the emission line while using the  parameters of the `L1' component taken from the first step. An example of the 
model can be seen in  Fig.~\ref{F:3Gfit}. The output of the fits is presented in Table~\ref{T:SCFb}.
We successively removed each component from the raw Doppler map  to probe the origin of the background. 
The result of the removal is shown in the middle and the bottom panels of Fig.~\ref{F:04}  where we removed the L1 component (middle panel) and both components (L1 + "spot") together (bottom panel). The last step of this procedure is not unambiguous. Nevertheless, the  Doppler map at the bottom panel clearly shows the presence of an extended ragged background. Its shape depends on the model parameters, however, in all cases,  we found that this background is concentrated around/close to  
the  WD position ($V_{\rm X} = 0$ km s$^{-1}$,
$V_{\rm Y} = -130$ km~s$^{-1}$) and has FWZI $\approx \pm600$km s$^{-1}$.
 There is also an additional stand-out excess of the flux in the background  at $V_{\rm X}\approx 250$ km~s$^{-1}$, $V_{\rm Y}\approx 0 $~km~s$^{-1}$. 
 We note that the maps do not show the usual `doughnut’ structure commonly observed from accretion disks in CVs.
 In addition, we constructed a Doppler map for \ion{He}{1}~6678 line in Fig.~\ref{F:04b}. The map shows only the emission from the L1 region.
 
 We interpret the obtained results as follows:
 \begin{enumerate}
 \item the `L1' component is formed at the irradiated side of the secondary close to the L1 point;
 \item the  `spot' component  ($V_{\rm X}\approx$ -300 km~s$^{-1}$, $V_{\rm Y}\approx$-100~km~s$^{-1}$) is located close to a standard position of the bright spot, however,  keeps out of Keplerian velocities of the disk or ballistic and Keplerian trajectories of the stream.  Thus we propose that this component is 
 more likely to be linked with radiation formed by a bright spot wind where, during a standstill, 
 our photometric modeling gives  the temperature 
 about 8200~K  (see Fig.~\ref{F:MCF1});
 \item the background is created mainly by the wind originating from the central hot part of the accretion disk and  partly  from an `outflow zone' of the accretion disk as it was observed in some NLs \citep{2010MmSAI..81..187B, 2017MNRAS.470.1960H, 2020MNRAS.497.1475S}. The presence of \ion{He}{2} in LAMOST spectra of AY~Psc also supports the wind origin of the H$\alpha$ line background by analogy with V~Sge-systems \citep{1998PASP..110..276S}.
 \end{enumerate}

\section{Discussion}
\label{sec:disc}

\begin{figure}
   \includegraphics{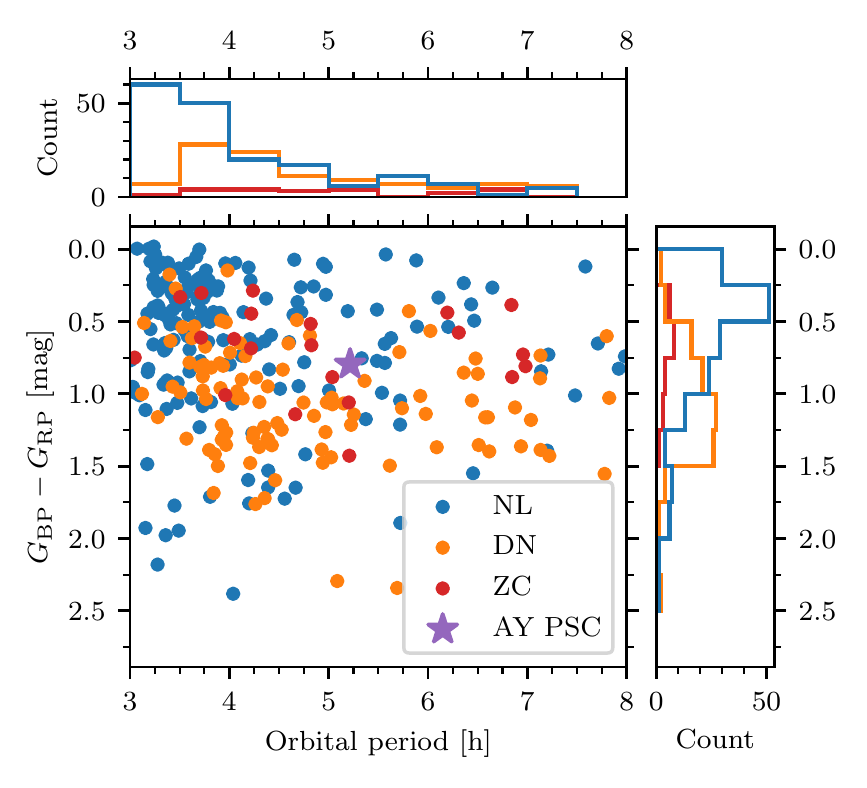}
   \includegraphics{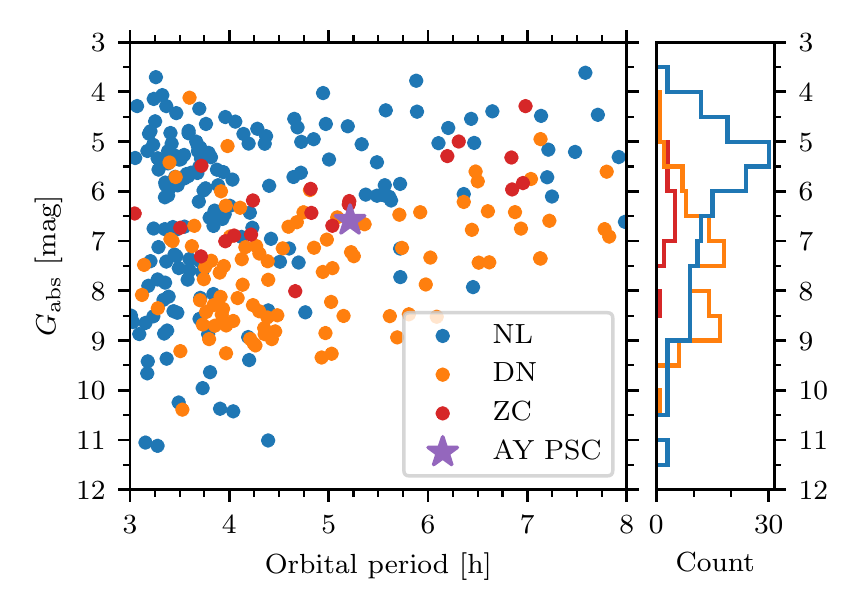}
         \caption{Different types of CVs displayed in orbital periods vs. {\it Gaia} color indexes (top) and absolute magnitudes  diagram. Orbital periods were taken from catalog of \cite{2003A&A...404..301R}, the color index was adopted from the {\it Gaia} catalog \citep[see][]{2016A&A...595A...1G, 2018A&A...616A...1G}.}
         \label{F:CVC}
  \end{figure}

\begin{figure}
   \centering
   
   \includegraphics{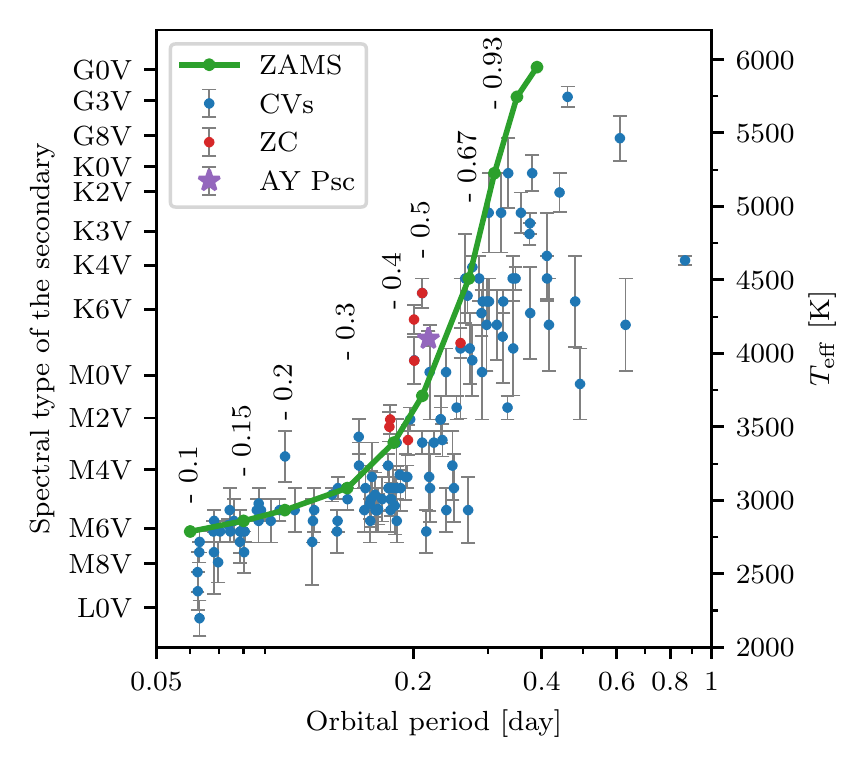}

         \caption{ Spectral types and effective temperatures of the donors of CVs (blue dots) plotted vs. their orbital periods
         (an update of the \citet{1998A&A...339..518B} list). The green line corresponds to the periods of semi-detached systems with main-sequence donors and 0.7~\ms\ accretor, the numbers mark the corresponding mass of main-sequence stars. The red dots show the Z~Cam systems listed in Table~\ref{tab:Ritter}, and AY~Psc is shown as a purple asterisk.   }
              
         \label{F:ZCamSp}
   \end{figure}

There are nineteen well-defined Z~Cam stars until now  \citep{2014JAVSO..42..177S}. They have orbital periods from about 3 to 8.5 hours with an average of 5.3 hours,  which is close to the orbital period of AY~Psc.
The mentioned period range is also filled by other types of non-magnetic CVs \citep{Warner:1995aa}  such as U~Gem-type DNe, VY~Scl-type, NLs SW~Sex, RW~Sex / RW~Tri types, and V~Sge-type systems. 
It is commonly accepted that the observational diversity of those types of CVs is caused by a difference in the mass transfer rate and, correspondingly,  by the state/temperature of the accretion disk. The last statement can be clearly seen  in  Fig.~\ref{F:CVC} where the {\sc Gaia} color index  $G_{\mathrm{BP}}-G_{\mathrm{RP}}$ (the top panel) and the absolute magnitude $G_{\mathrm{abs}}$ (the bottom panel) versus orbital periods are shown.  The NLs look brighter $G_{\rm abs, \; NL}=5.0(5)$ and bluer $(G_{\rm BP} - G_{\rm RP})_{\rm NL}\sim 0.3(2)$ in comparison with DNe ($G_{\rm abs, \; DNe}=8(1)$,  $(G_{\rm BP} - G_{\rm RP})_{\rm DNe}\sim 1.3(2)$. 
The Z~Cam-type objects fill the space between them with $G_{\rm abs,~Z~Cam}=6.5(5)$,  $(G_{\rm BP} - G_{\rm RP})_{\rm ~Z~Cam}\sim 0.7(2)$.  A high dispersion in color indices and absolute magnitudes is caused by the absence of exact values of inclinations of most of the systems (see the right histograms of both panels). 
Nevertheless, the indicated trend is clearly visible.
It is also interesting that eclipsing NLs with known parameters (V347~Pup, RW~Tri, UX~UMa, 1RXS J064434.5+334451) in the period range of 4--5 hours have slightly less massive primaries ($\sim 0.7$ \ms) and a slightly hotter ($\sim$M0V/3900K$\pm$400K) secondaries in comparison with DNe (U~Gem, CW~Mon, GY~Cnc, and others $\sim 1.0$~\ms, $\sim$M3.5V/3300K$\pm$270K) \citep[see data and the latest references for these objects therein\footnote{\url{http://simbad.u-strasbg.fr/simbad/}}]{2003A&A...404..301R, 2000A&AS..143....9W}. The average secondary mass is about 0.4~\ms\ in both cases. The estimate of a spectral class of the secondary in Z~Cam systems looks similar to NLs with a large dispersion (M0[$\pm3$]V, see Table~\ref{tab:Ritter}).  Nevertheless, they appear hotter than secondaries  in most CVs of the same orbital periods (Fig.~\ref{F:ZCamSp}).  
Unfortunately, until now, the number of systems with well-determined fundamental parameters is low, and the above-mentioned tendency must be fully confirmed.

In the list of confirmed Z~Cam systems, only AY~Psc shows deep eclipses that, together with the time-resolved spectroscopic data and the {\sc Gaia} distance, provide a possibility to determine the system parameters.   
 From the light curve fitting of AY~Psc we found that the mass of the WD primary is 
$M_{\rm WD}=0.90(4)$ M$_\sun$, the mass ratio is $q=0.50(4)$, and the system inclination is $i=$ 74\fdg8(7).   At the practically same inclination angle, we obtained less massive components in the system comparison with masses presented by \citet{1993AJ....106..311H} whose estimations were based on the analytical technique applied for analysis of the eclipse profile (see references therein).  Our system parameters exclude AY~Psc as a  potential SN~Ia progenitor. 
The WD primary has a temperature of about 30~000~K, and the red dwarf temperature is higher than can be expected for the secondaries from the orbital period vs. secondary spectral class relation \citep{2011ApJS..194...28K}.
Nevertheless, the characteristics of  AY~Psc are very similar to other Z~Cam-systems with close orbital periods, such as  
AT~Cnc \citep{1999PASJ...51..115N} and 
RX~And \citep{2002ApJ...574..937S, 2001ApJ...555..834S}.

We also found that in the low state, the eclipsing light curve can be well described by a steady-state disk with an accretion rate of 2.4$\times$10$^{-10}$ M$_\sun$ year$^{-1}$. The object out-of-eclipse magnitude shows dispersion about $\sim$0.5~mag 
in the low brightness state. Also, the disk is close to the instability condition and goes into an outburst  at a relatively short time scale. This means that the estimation of the mass transfer rate from the analysis of the particular eclipse light curve gives only the order of magnitude of $\dot{M}$ value for quiescence. However, similar  accretion rates in quiescence were found for other Z~Cam systems, e.g., RX~And \citep{2002ApJ...574..937S} and EM~Cyg \citep{2010PASJ...62..965D}. 
The outer part of the quiescence disk has an effective temperature in the range of $\sim$3000--5000~K, and its brightness distribution looks like in a steady-state disk. 

The disk in outburst and standstill is significantly hotter. The effective temperature departs from the steady-state disk dependence, and the optical depth, where the continuum is formed, extends in the z-direction especially more significantly in the outburst maximum. 
The \ion{He}{2} emission $\lambda$4686 in outburst spectra supports the idea of forming of wind from the disk in outbursts and standstill.
It is also the distinguishing feature of high-mass transfer ($>10^{-9}$ M$_\odot$ year$^{-1}$) V~Sge-type systems \citep{1998PASP..110..276S}  which represent a  state of binary evolution dominated by wind mass loss. The example of such an object with an orbital period close to AY~Psc is V617~Sgr \citep{1999AJ....117..534C} in which a strong wind from the stream/disk impact region is proposed \citep[see fig.~6 therein]{1999A&A...351.1021S}. 
The effective temperature of the primary more probably varies with the change of the system state as it was found, for example,
in  RX~And \citep{2002ApJ...574..937S}. Still, its contribution to optical flux in an outburst or standstill 
is negligible compared to the flux from the inner part of the disk. Thus, using the constant value for the light curve fitting in the outburst and standstill does not affect obtained results.

\begin{table}[!tbp]
    \caption{List of known Z~Cam-type CVs with the orbital period between 0.17 -- 0.26~day and estimate of the secondary star type \citep{2003A&A...404..301R}. }
    \begin{center}
    \begin{tabular}{ccccc}
    \hline
    System        &    $V$ [mag] &  Period [day] & Sp. Type & Ref.\\
  \hline
       WW Cet     &    13.9   & 0.175800  & M2.5V  & 1  \\
        ES Dra    &    16.2   & 0.176600  & M1-3V  & 2  \\
     HS2325+8205 *  &    16.4   & 0.194335  & M2-4V  & 3 \\  
        HX Peg    &    15.8   & 0.200800  & K6     & 4 \\         
        AT Cnc    &    15.3   & 0.201100  & K7-M0V & 5  \\        
        RX And    &    14.0   & 0.209893  & K5V    & 6  \\                     
        AY Psc    &    15.3   & 0.217321  & K7V    & 7  \\            
         AH Her   &    13.9   & 0.258116  & K7V    & 8  \\        
    \hline\noalign{\smallskip}
    \end{tabular}
    \end{center}
   \tablecomments{ 
                1 - \cite{2006MNRAS.368..361P},  
               2 - \cite{2012NewA...17..108R},
               3 - \cite{2012PASP..124..204P},
               4 - \cite{1994MNRAS.270..804R},
               5 - \cite{2012ApJ...758..121S},
               6 - \cite{2006MNRAS.373..484K},
               7 - this paper,
               8 - \cite{2002MNRAS.337.1215N}. \\
               {*}{Described by \cite{2012PASP..124..204P} as a plausible Z~Cam-type candidate.}      }                        
    \label{tab:Ritter}
  
\end{table}

 The Doppler map and emission line trailed spectra of AY~Psc in standstill are similar to those observed in other NLs, e.g. LX~Ser \citep{1994ApJS...93..519K},
 RW~Sex \citep{2017MNRAS.470.1960H},
 1RXS~J064434.5+334451 \citep{2017MNRAS.470.1960H}, 
 V347~Pup \citep{ 2005MNRAS.357..881T}, 
 V341~Ara \citep{2021MNRAS.501.1951C}, 
 BP~Lyn \citep{1996MNRAS.282..943S}, 
 HBHA~4705-03 \citep{2013AstL...39...38Y}, 
 AC~Cnc \citep{2004MNRAS.353.1135T}, 
 V363~Aur\citep{1994ApJS...93..519K},
 RW Tri \citep{2020MNRAS.497.1475S},
 and some DNe, e.g. EM~Cyg \citep{2000MNRAS.313..383N}, AH~Her, SS~Cyg \citep{2002MNRAS.337.1215N}, and CzeV404~Her \citep{2021A&A...652A..49K}. While in most of them, the low-resolution spectra show only the one-peaked Balmer 
 emission lines, the higher-resolution spectroscopy reveals a structure with two or more components. 
 Using the terminology proposed by \citet{2017MNRAS.470.1960H} for NLs,   the  `narrow' low-velocity component, if is present, usually originates from the disk-illuminated side of the secondary and is identical to the `L1' component here. Other wider component(s) have not been related to  Keplerian motions in the accretion disk and, more probably, are formed by winds from the bright spot and/or the central part of the accretion disk and  from 
 the outflow zone of the disk \citep{2017MNRAS.470.1960H, 2020MNRAS.497.1475S, 2021MNRAS.503.1431H}. 
 The specific view of the Doppler map with `wide' component(s) depends on the physical conditions where they are formed. It is a very interesting subject, but its full discussion is out of the scope of the presented paper.
 A `doughnut' structure associated with the Balmer emission from the accretion disk is absent in all objects mentioned in this section. 
 
 The interesting feature of some long orbital period CVs is the relation between the depth of an eclipse and out-of-eclipse brightness of objects \citep[and references therein]{2020MNRAS.497.1475S}. We showed that AY~Psc also suits this relation. The proposed model can be easily explained by the change in temperature distribution followed by the variation of the disk height along its radius. Nonetheless, it cannot be excluded that some additional effects as tidal torque by the secondary on a misaligned accretion disk are also possible
\citep{2009ApJ...705..603M}.

\section{Conclusions}
\label{sec:concl}

We performed time-resolved photometric and spectroscopic observations of the eclipsing Z~Cam-type 
CV AY~Psc.
Based on the fitting of the eclipse light curves  in different states of the system and time-resolved spectroscopy obtained in a standstill state, we determined:
\begin{enumerate}

\item The system contains a hot massive white dwarf $M_{\mathrm WD} = 0.90(4)$ M$_\sun$.  
The mass of the secondary is $M_2=0.45(5)$~\ms. However, its
effective temperature is $T_2 = 4100$(50)~K.  It corresponds to the spectral type of K7V. 
We note that it is significantly hotter than the effective temperature of the main-sequence star with the corresponding mass whose radius must also be less than 25\% size of the secondary. The last means that the secondary is an evolved star with a flayed shell. The system inclination is $i=74\fdg8(7)$.

\item The mass accretion rate and the disk structure change between quiescence, standstill, and outburst. The mass accretion rate increases from quiescence to  outburst maximum, and the disk becomes hotter and its shape flatter.  

\item The H$\alpha$ emission line  shows a multi-component structure similar to those observed in other long orbital period NLs and DNe. We conclude that the H$\alpha$ emission line is formed by the superposition of  radiation from the irradiated surface of the secondary, the wind from the bright spot and hot central part of the accretion disk, and from the outflow zone.

\item We found that AY~Psc shows two distinct types of outbursts during its cycles, which differ in the shape of their light curve and their duration:  long and short outbursts. 
 The long outbursts  exhibit a plateau phase after reaching their peak brightness, which is a shape typical for superoutbursts of SU~UMa-type systems and long outbursts of U~Gem and Z~Cam CVs \citep[see][fig.~1]{2016MNRAS.460.2526O}.  Individual long outbursts  differ in the duration of the plateau phase and subsequent decline rate. Short outbursts do not exhibit a plateau phase, and they have a shape typical for normal outbursts of DNe. 
Long outbursts were detected more prominently, and no pattern in the occurrence of any particular outburst type was found in the available data. The prominence of long outbursts over short outbursts is unusual, as typically long outbursts have longer occurrence cycles than short outbursts \citep{2016MNRAS.460.2526O}.

\item We calculated a new ephemeris of AY~Psc using data covering over 30 years of observations. 
Instead of a simple linear ephemeris fit proposed by previous studies, the updated $O-C$ diagram  shows a quadratic ephemeris fit corresponding to the increase of the orbital period  with a rate of  $dP/dt = {+ 7.6}\times10^{-9} \mathrm{\; d \; year}^{-1}$. 
Nevertheless, even when we take the quadratic ephemeris into consideration, the $O-C$ diagram shows a large scatter. This can be caused by a variation of mid-eclipse times due to a change of the accretion disk radius and/or azimuthal movements of the bright spot position at the edge of the accretion disk \citep{2021RNAAS...5..148S}. 
The increase of the orbital period could be an effect of the dynamical evolution of the system, namely the effect of changes on the masses of the primary and the secondary due to mass transfer.
Another possibility is an effect of a third body in the system that cannot be ruled out as the observed parabolic behavior could be caused by a cyclical change of the period longer than the time-span of current observations \citep[see][for examples of CVs with proposed third bodies]{2020RMxAA..56...19C, 2020ApJ...901..113F, 2021MNRAS.505..677L}.

\end{enumerate}

\noindent
We think that this object is a unique laboratory to study the response of the accretion disk on mass transfer variations. 
New precise time-resolved simultaneous photometric and spectroscopic observations during active and standstill states are needed in the future to understand the mechanism of transition between them better.

\begin{acknowledgements}

We are grateful to the anonymous referee for the useful and valuable suggestions that allowed us to improve the manuscript.
The authors thank Prof.~H.~G\"{u}lsecen, Istanbul University, Turkey, who kindly provided us with his original photometry of AY~Psc. 
Based upon observations carried out at the Observatorio Astron\'{o}mico Nacional on the Sierra San Pedro M\'{a}rtir (OAN SPM), Baja California, México.
S.Z. acknowledges PAPIIT grants IN102120 and IN119323.
The research of J.K., M.W., and J.M. was also supported by the project {\sc Cooperatio - Physics} of Charles University in Prague. This research has been funded in a part by the
Science Committee of the Ministry of Science and Higher Education of the Republic of Kazakhstan (grant No.  AP19678376  ). S.K. acknowledges Al-Farabi Kazakh National University for a postdoc position. 
This work has made use of data from the European Space Agency (ESA) mission 
{\it Gaia} (\url{https://www.cosmos.esa.int/gaia}), processed by the Gaia Data Processing and Analysis Consortium 
(DPAC, \url{https://www. cosmos.esa.int/web/gaia/dpac/consortium}). 
Funding for the DPAC has been provided by national institutions, 
in particular, the institutions participating in the Gaia Multilateral Agreement.
We acknowledge with thanks the variable star observations from the AAVSO International Database contributed by observers worldwide and used in this research.
The following internet-based resources were used in research for this paper:
the SIMBAD database and the VizieR service operated at CDS, Strasbourg, France; NASA's Astrophysics Data System Bibliographic Services.
This paper includes data collected by the TESS mission. Funding for the TESS mission is provided by NASA's Science Mission Directorate.
The TESS data presented in this paper were obtained from the Mikulski Archive for Space Telescopes (MAST) at the Space Telescope Science Institute. The specific observations analyzed can be accessed via \dataset[https://doi.org/10.17909/5fq3-s930]{https://doi.org/10.17909/5fq3-s930}. STScI is operated by the Association of Universities for Research in Astronomy, Inc., under NASA contract NAS5–26555. Support to MAST for these data is provided by the NASA Office of Space Science via grant NAG5–7584 and by other grants and contracts.
Based on observations obtained with the Samuel Oschin 48-inch Telescope at the Palomar Observatory as part of the Zwicky Transient Facility project. ZTF is supported by the National Science Foundation under Grant No. AST-1440341 and a collaboration including Caltech, IPAC, the Weizmann Institute for Science, the Oskar Klein Center at Stockholm University, the University of Maryland, the University of Washington, Deutsches Elektronen-Synchrotron and Humboldt University, Los Alamos National Laboratories, the TANGO Consortium of Taiwan, the University of Wisconsin at Milwaukee, and Lawrence Berkeley National Laboratories. Operations are conducted by the COO, IPAC, and UW.
\end{acknowledgements}

\software{  Matplotlib \citep{Hunter:2007},
            NumPy \citep{harris2020array},
            SciPy \citep{2020SciPy-NMeth},
            astropy \citep{2013A&A...558A..33A,2018AJ....156..123A, 2022ApJ...935..167A},
            Fast maximum entropy Doppler mapping \citep{1998astro.ph..6141S},
            PyDoppler \citep{2021ascl.soft06003H},
            IRAF \citep{1986SPIE..627..733T, 1993ASPC...52..173T},
            CVlab \citep{2013A&A...549A..77Z} 
          }

\bibliographystyle{aasjournal.bst}
\bibliography{bibliography.bib}

\end{document}